\RequirePackage{fix-cm}
\PassOptionsToPackage{square,numbers,sort&compress}{natbib}
\documentclass[natbib,twocolumn]{svjour3_arxiv} 
\bibpunct[,]{[}{]}{,}{n}{,}{,} 
\smartqed  
\usepackage{cancel}
\usepackage{graphicx}
\usepackage{amsmath,amssymb,amsfonts}
\usepackage{graphicx}
\usepackage{epsfig}
\usepackage{xcolor}
\usepackage{fancyvrb}
\usepackage{stmaryrd}
\usepackage{upgreek}
\usepackage{caption}
\usepackage{color, colortbl}
\graphicspath{./}

\usepackage{hyperref}
\makeatletter
\hypersetup{
    colorlinks,
    linkcolor={blue!50!black},
    citecolor={blue!50!black},
    urlcolor={blue!80!black}
}
\AtBeginDocument{%
  \setlength{\oddsidemargin}{\dimexpr(\paperwidth-\textwidth)/2-1in}%
  \setlength{\evensidemargin}{\oddsidemargin}%
  \setlength{\topmargin}{%
    \dimexpr(\paperheight-\textheight)/2-\headheight-\headsep-1in}%
}
\makeatother

\usepackage{graphicx}
\usepackage{multirow}      
\usepackage{wasysym}
\newcommand{\EE}{\begin{equation}}
\newcommand{\Ee}{\end{equation}}
\newcommand{\FF}[1]{\begin{figure}[#1]\CT}
\newcommand{\Ff}{\end{figure}}

\newcommand{\TT}[1]{\begin{table}[#1]}
\newcommand{\Tt}{\end{table}}

\newcommand{\TB}[1]{\begin{tabular}{#1}}
\newcommand{\Tb}{\end{tabular}}
\newcommand{\II}[1]{\begin{itemize}[#1]}
\newcommand{\Ii}{\end{itemize}}
\newcommand{\EN}[1]{\begin{enumerate}[#1]}
\newcommand{\En}{\end{enumerate}}

\newcommand{\CT}{\centering}
\newcommand{\gaq}{\gamma_{\rm eq}}

\newcommand{\red}{\color{black}}

\newcommand\rtwoC[1]{\textcolor{black}{#1}}
\definecolor{mulberry}{rgb}{0.77, 0.29, 0.55}
\newcommand{\revC}[1]{{\color{black}#1}}
\newcommand{\mw}[1]{{\color{black}#1}}
\newcommand{\mww}[1]{{\color{mulberry}#1}}
\newcommand{\revM}[1]{{\color{red}#1}}
\renewcommand{\mww}[1]{{\color{black}#1}}
\renewcommand{\revM}[1]{{\color{black}#1}}

\newcommand{\figref}[1]{Fig.~\ref{#1}}

\newcommand{\tabreff}[1]{Table~\ref{#1}}
\newcommand{\cip}[1]{\citep{#1}}

\newcommand{\secref}[1]{Section~\ref{#1}}

\graphicspath{{figures/}{Figures/}}

\begin{document}\sloppy
%
%
\newcommand{\Ds}{\displaystyle}
\newcommand{\Ts}{\textstyle}
\newcommand{\Ss}{\scriptstyle}%
%
%
\renewcommand{\d}{{\,\rm  d}}
\newcommand{\D}{{\,\rm  D}}
\newcommand{\dt}{\triangle t}
\newcommand{\T}[1]{{#1}^{\sf  T}}
\newcommand{\I}[1]{{#1}^{\rm -1}}
\newcommand{\IT}[1]{{#1}^{\sf  -T}}
\newcommand{\pd}[2]{\displaystyle\frac{\partial #1}{\partial #2}}
\newcommand{\pds}[2]{\displaystyle\frac{\partial\,#1}{\partial #2}}
\newcommand{\td}[2]{\frac{{\rm d} #1}{{\rm d} #2}}
\newcommand{\pdf}[2]{\displaystyle{\partial #1}{/\partial #2}}
\newcommand{\tdf}[2]{\displaystyle{\d #1}{/\d #2}}
\newcommand{\z}[1]{{\tilde{#1}}}
\newcommand{\abs}[1]{|#1|}
\newcommand{\absg}[1]{\left| #1 \right|}
\newcommand{\norm}[1]{\Vert #1 \Vert}
\newcommand{\normg}[1]{\left\Vert \, #1 \, \right\Vert}
\newcommand{\sign}[1]{{\rm sgn}\left( #1 \right)}
\newcommand{\grad}[1]{{\rm grad}\left( #1 \right)}
\renewcommand{\div}[1]{{\rm div }\left( #1 \right)}
\newcommand{\rot}[1]{{\rm curl }\left( #1 \right)}
\newcommand{\Grad}[1]{{\rm Grad}\left( #1 \right)}
\newcommand{\Div}[1]{{\rm Div }\left( #1 \right)}
\newcommand{\Rot}[1]{{\rm Curl }\left( #1 \right)}
\newcommand{\tr}[1]{{\rm sp  }( #1 )}
\newcommand{\cof}[1]{{\rm cof  }( #1 )}
\renewcommand{\det}[1]{{\rm det }( #1 )}
\newcommand{\dev}[1]{{\rm dev}( #1 )}
\newcommand{\fsym}[1]{{\rm sym }( #1 )}
\newcommand{\fskw}[1]{{\rm skw }( #1 )}
\newcommand{\unit}[1]{{\rm #1}}
\newcommand{\lb}{\left(}
\newcommand{\rb}{\right)}
\newcommand{\la}{\langle}
\newcommand{\ra}{\rangle}
\newcommand{\ls}{\{}
\newcommand{\rs}{\}}
%
%
\newcommand{\cred }{\color{red}}
\newcommand{\cblue }{\color{blue}}
\newcommand{\cgreen }{\color{green}}
%
%
\newcommand{\ma}{$\bigstar$}
\newcommand{\mb}{$\blacksquare\;$}
%
%
\newcommand{\fempty}[1]{{}}
%
%
\newcommand{\f}[1]{\mbox{$ #1 $}}
\newcommand{\topic}[1]{\\[0.8ex]{\bf  #1.}}
\newcommand{\topics}[1]{\vspace*{0.8ex}{\bf  #1.}}
%
%
\newcommand{\ta}{{}^n}
\newcommand{\tb}{{}^{n+1}}
\newcommand{\tba}{{}^{n+1}_i}
\newcommand{\tbb}{{}^{n+1}_{i+1}}
\renewcommand{\r}[2]{{#2_{\langle #1 \rangle}}}
%
%
\newcommand{\vect} {{\it Vect}}
\newcommand{\ortp} {{\it Orth}^{\small +}}
\newcommand{\ort} {{\it Orth}}
\newcommand{\sym} {{\it Sym}}
\newcommand{\skw} {{\it Skw}}
\newcommand{\psym}{{\it Psym}}
\newcommand{\unim}{{\it Unim}}
\newcommand{\invp}{{\it Inv}^{\small +}}
\newcommand{\inv}{{\it Inv}}
\newcommand{\reell}{{\it R}}
\newcommand{\reellp}{{\it R}^+}
\newcommand{\lin} {{\it Lin}}
\newcommand{\irr} {{\it Sym'}}
\newcommand{\sop} {{{SO}(3)}}
%
%
\newcommand{\rI}{{\mathit{I}}}
\newcommand{\rII}{{\mathit{II}}}
\newcommand{\rIII}{{\mathit{III}}}
\newcommand{\rIV}{{\mathit{IV}}}
%
%
\newcommand{\sty}[1]{\mbox{\boldmath $#1$}}
\newcommand{\styy}[1]{{\mathbb{#1}}}
\newcommand{\fa}{\sty{ a}}
\newcommand{\fb}{\sty{ b}}
\newcommand{\fc}{\sty{ c}}
\newcommand{\fd}{\sty{ d}}
\newcommand{\fe}{\sty{ e}}
\newcommand{\ff}{\sty{ f}}
\newcommand{\fg}{\sty{ g}}
\newcommand{\fh}{\sty{ h}}
\newcommand{\fj}{\sty{ j}}
\newcommand{\fk}{\sty{ k}}
\newcommand{\fl}{\sty{ l}}
\newcommand{\fm}{\sty{ m}}
\newcommand{\fn}{\sty{ n}}
\newcommand{\fo}{\sty{ o}}
\newcommand{\fp}{\sty{ p}}
\newcommand{\fq}{\sty{ q}}
\newcommand{\fr}{\sty{ r}}
\newcommand{\fs}{\sty{ s}}
\newcommand{\ft}{\sty{ t}}
\newcommand{\fu}{\sty{ u}}
\newcommand{\fv}{\sty{ v}}
\newcommand{\fw}{\sty{ w}}
\newcommand{\fx}{\sty{ x}}
\newcommand{\fy}{\sty{ y}}
\newcommand{\fz}{\sty{ z}}
\newcommand{\fzero}{\sty{ 0}}
\newcommand{\cfu}{\check{\fu}}
\newcommand{\cfv}{\check{\fv}}
\newcommand{\cfp}{\check{\fp}}
\newcommand{\cfD}{\check{\fD}}
\newcommand{\cfeps}{\check{\feps}}
\newcommand{\cftau}{\check{\ftau}}
\newcommand{\cfsigma}{\check{\fsigma}}
\newcommand{\hfu}{\hat{\fu}}
\newcommand{\hfeps}{\hat{\feps}}
\newcommand{\hfsigma}{\hat{\fsigma}}
\newcommand{\ffa}{\styy{ a}}
\newcommand{\ffb}{\styy{ b}}
\newcommand{\ffc}{\styy{ c}}
\newcommand{\ffd}{\styy{ d}}
\newcommand{\ffe}{\styy{ e}}
\newcommand{\fff}{\styy{ f}}
\newcommand{\ffg}{\styy{ g}}
\newcommand{\ffh}{\styy{ h}}
\newcommand{\ffj}{\styy{ j}}
\newcommand{\ffk}{\styy{ k}}
\newcommand{\ffl}{\styy{ l}}
\newcommand{\ffm}{\styy{ m}}
\newcommand{\ffn}{\styy{ n}}
\newcommand{\ffo}{\styy{ o}}
\newcommand{\ffp}{\styy{ p}}
\newcommand{\ffq}{\styy{ q}}
\newcommand{\ffr}{\styy{ r}}
\newcommand{\ffs}{\styy{ s}}
\newcommand{\fft}{\styy{ t}}
\newcommand{\ffu}{\styy{ u}}
\newcommand{\ffv}{\styy{ v}}
\newcommand{\ffw}{\styy{ w}}
\newcommand{\ffx}{\styy{ x}}
\newcommand{\ffy}{\styy{ y}}
\newcommand{\ffz}{\styy{ z}}
\newcommand{\fA}{\sty{ A}}
\newcommand{\fB}{\sty{ B}}
\newcommand{\fC}{\sty{ C}}
\newcommand{\fD}{\sty{ D}}
\newcommand{\fE}{\sty{ E}}
\newcommand{\fF}{\sty{ F}}
\newcommand{\zfF}{\tilde{\sty{ F}}}
\newcommand{\fG}{\sty{ G}}
\newcommand{\fH}{\sty{ H}}
\newcommand{\fI}{\sty{ I}}
\newcommand{\fJ}{\sty{ J}}
\newcommand{\fK}{\sty{ K}}
\newcommand{\fL}{\sty{ L}}
\newcommand{\fM}{\sty{ M}}
\newcommand{\fN}{\sty{ N}}
\newcommand{\fO}{\sty{ 0}}
\newcommand{\fP}{\sty{ P}}
\newcommand{\fQ}{\sty{ Q}}
\newcommand{\fR}{\sty{ R}}
\newcommand{\fS}{\sty{ S}}
\newcommand{\fT}{\sty{ T}}
\newcommand{\fU}{\sty{ U}}
\newcommand{\fV}{\sty{ V}}
\newcommand{\fW}{\sty{ W}}
\newcommand{\fX}{\sty{ X}}
\newcommand{\fY}{\sty{ Y}}
\newcommand{\fZ}{\sty{ Z}}
\newcommand{\bfA}{\bar{\sty{ A}}}
\newcommand{\bfB}{\bar{\sty{ B}}}
\newcommand{\bfC}{\bar{\sty{ C}}}
\newcommand{\bfD}{\bar{\sty{ D}}}
\newcommand{\bfE}{\bar{\sty{ E}}}
\newcommand{\bfF}{\bar{\sty{ F}}}
\newcommand{\bfG}{\bar{\sty{ G}}}
\newcommand{\bfH}{\bar{\sty{ H}}}
\newcommand{\bfI}{\bar{\sty{ I}}}
\newcommand{\bfJ}{\bar{\sty{ J}}}
\newcommand{\bfK}{\bar{\sty{ K}}}
\newcommand{\bfL}{\bar{\sty{ L}}}
\newcommand{\bfM}{\bar{\sty{ M}}}
\newcommand{\bfN}{\bar{\sty{ N}}}
\newcommand{\bfO}{\bar{\sty{ 0}}}
\newcommand{\bfP}{\bar{\sty{ P}}}
\newcommand{\bfQ}{\bar{\sty{ Q}}}
\newcommand{\bfR}{\bar{\sty{ R}}}
\newcommand{\bfS}{\bar{\sty{ S}}}
\newcommand{\bfT}{\bar{\sty{ T}}}
\newcommand{\bfU}{\bar{\sty{ U}}}
\newcommand{\bfV}{\bar{\sty{ V}}}
\newcommand{\bfW}{\bar{\sty{ W}}}
\newcommand{\bfX}{\bar{\sty{ X}}}
\newcommand{\bfY}{\bar{\sty{ Y}}}
\newcommand{\bfZ}{\bar{\sty{ Z}}}
\newcommand{\bfeps}{\bar{\feps}}
\newcommand{\bfsigma}{\bar{\fsigma}}
\newcommand{\bftau}{\bar{\ftau}}
\newcommand{\ffA}{\styy{ A}}
\newcommand{\ffB}{\styy{ B}}
\newcommand{\ffC}{\styy{ C}}
\newcommand{\ffD}{\styy{ D}}
\newcommand{\ffE}{\styy{ E}}
\newcommand{\ffF}{\styy{ F}}
\newcommand{\ffG}{\styy{ G}}
\newcommand{\ffH}{\styy{ H}}
\newcommand{\ffI}{\styy{ I}}
\newcommand{\ffJ}{\styy{ J}}
\newcommand{\ffK}{\styy{ K}}
\newcommand{\ffL}{\styy{ L}}
\newcommand{\ffM}{\styy{ M}}
\newcommand{\ffN}{\styy{ N}}
\newcommand{\ffO}{\styy{ O}}
\newcommand{\ffP}{\styy{ P}}
\newcommand{\ffQ}{\styy{ Q}}
\newcommand{\ffR}{\styy{ R}}
\newcommand{\ffS}{\styy{ S}}
\newcommand{\ffT}{\styy{ T}}
\newcommand{\ffU}{\styy{ U}}
\newcommand{\ffV}{\styy{ V}}
\newcommand{\ffW}{\styy{ W}}
\newcommand{\ffX}{\styy{ X}}
\newcommand{\ffY}{\styy{ Y}}
\newcommand{\ffZ}{\styy{ Z}}
\newcommand{\falpha}{\mbox{\boldmath $\alpha$}}
\newcommand{\fbeta}{\mbox{\boldmath $\beta$}}
\newcommand{\fchi}{\mbox{\boldmath $\chi$}}
\newcommand{\ftau}{\mbox{\boldmath $\tau$}}
\newcommand{\fsigma}{\mbox{\boldmath $\sigma$}}
\newcommand{\fDelta}{\mbox{\boldmath $\Delta$}}
\newcommand{\fomega}{\mbox{\boldmath $\omega$}}
\newcommand{\fOmega}{\mbox{\boldmath $\Omega$}}
\newcommand{\fLambda}{\mbox{\boldmath $\Lambda$}}
\newcommand{\fxi}{\mbox{\boldmath $\xi $}}
\newcommand{\fXi}{\mbox{\boldmath $\Xi $}}
\newcommand{\feps}{\mbox{\boldmath $\varepsilon $}}
\newcommand{\fgamma}{\mbox{\boldmath $\gamma $}}
\newcommand{\fGamma}{\mbox{\boldmath $\Gamma $}}
\newcommand{\fSigma}{\mbox{\boldmath $\Sigma $}}
\newcommand{\fPi}{\mbox{\boldmath $\Pi $}}
\newcommand{\fPhi}{\mbox{\boldmath $\Phi $}}
\newcommand{\fphi}{\mbox{\boldmath $\phi $}}
\newcommand{\fvarphi}{\mbox{\boldmath $\varphi $}}
\newcommand{\fPsi}{\mbox{\boldmath $\Psi $}}
\newcommand{\fJota}{{\bf I}}
\newcommand{\fepsi}{\mbox{\boldmath $\varepsilon $}}
\newcommand{\cA}{{\cal A}}
\newcommand{\cB}{{\cal B}}
\newcommand{\cC}{{\cal C}}
\newcommand{\cD}{{\cal D}}
\newcommand{\cE}{{\cal E}}
\newcommand{\cF}{{\cal F}}
\newcommand{\cG}{{\cal G}}
\newcommand{\cH}{{\cal H}}
\newcommand{\cI}{{\cal I}}
\newcommand{\cJ}{{\cal J}}
\newcommand{\cK}{{\cal K}}
\newcommand{\cL}{{\cal L}}
\newcommand{\cM}{{\cal M}}
\newcommand{\cN}{{\cal N}}
\newcommand{\cO}{{\cal O}}
\newcommand{\cP}{{\cal P}}
\newcommand{\cp}{{\cal p}}
\newcommand{\cQ}{{\cal Q}}
\newcommand{\cR}{{\cal R}}
\newcommand{\cS}{{\cal S}}
\newcommand{\cT}{{\cal T}}
\newcommand{\cU}{{\cal U}}
\newcommand{\cV}{{\cal V}}
\newcommand{\cW}{{\cal W}}
\newcommand{\cX}{{\cal X}}
\newcommand{\cY}{{\cal Y}}
\newcommand{\cZ}{{\cal Z}}

\title{Modeling contrary size effects of tensile- and torsion-loaded oligocrystalline gold microwires
}


\author{E.~Bayerschen \and A.~Prahs \and S.~Wulfinghoff \and M.~Ziemann \and P.A.~Gruber \and M.~Walter \and T.~B\"ohlke 
}


\institute{E.~Bayerschen \and A.~Prahs \and T.~B\"ohlke \at
              Institute of Engineering Mechanics (ITM), Chair for Continuum Mechanics, Karlsruhe Institute of Technology (KIT), Kaiserstr.\ 10, D-76131 Karlsruhe, Germany \\
              \email{eric.bayerschen@kit.edu, andreas.prahs@kit.edu, thomas.boehlke@kit.edu}\\
              Tel.: +49-721-608-488-52
           \and
           S.~Wulfinghoff \at
              RWTH Aachen University, Institute of Applied Mechanics, Mies-van-der-Rohe-Str. 1, D-52074 Aachen, Germany\\
\email{stephan.wulfinghoff@rwth-aachen.de}
\and
M.Ziemann  \and P.A.~Gruber \and M.~Walter \at Institute for Applied Materials (IAM), Karlsruhe Institute of Technology (KIT), Engelbert-Arnold-Strasse 4, D-76131 Karlsruhe, Germany\\
\email{michael.ziemann@kit.edu, mario.walter@kit.edu, patric.gruber@kit.edu}
}

\date{Received: date / Accepted: date}

{\date{The published version of this article is available at Springer in the \href{http://dx.doi.org/10.1007/s10853-016-0020-7}{Journal of Materials Science}.}}
\maketitle

\begin{abstract}
\mw{When Chen et al.\ (2015, {\it Acta Mater.} {\bf 87}, 78-85) investigated the deformation behavior of oligocrystalline gold microwires with varying diameters in both uniaxial tension and torsion, contrary size effects were observed for the different load cases. In accompanying microstructural studies it was found, that the microwires of different thickness reveal \mww{distinctive} differences in grain size and texture, respectively. As a consequence, a significant influence of these microstructural variations on the determined size effects was assumed. However, within the frame of their work, a direct confirmation could only be presented for the effect of the grain size.} In the present work, the size-dependent mechanical response of the microwires is modeled with a gradient plasticity theory. By finite element simulations of simplified grain aggregates, the influence of the texture on the size effects is investigated under both loading conditions. \mw{It is shown that the experimentally observed contrary size effects can only be reproduced when taking into account the individual textures of the microwires of different thickness within the modeling.}
\end{abstract}

\section{Introduction}
Size effects in the mechanical response of metallic microspecimens are observed under various load settings, e.g., for torsion of thin wires~\cite{fleck1994strain}, indentation~\cite{ma1995size}, or bending of thin foils~\cite{stolken1998microbend}. \mw{The common finding is that strength is increasing with decreasing size.} It has been determined that both the microstructure of the material - in specific the grain size - and the overall specimen size influence such size effects~\cite{yang2012yield}. Grain boundaries (GBs) contribute significantly to this effect due to dislocations piling up in their vicinity. Dislocations piling up at GBs is a pronounced mechanism in polycrystalline and in oligocrystalline microstructured materials. Therefore, the modeling of GBs and their role as obstacles for dislocation movement is of a central interest in the development of plasticity models. The influence of dislocations results from both statistically stored dislocations (SSDs) and geometrically necessary dislocations (GNDs), e.g.,~\cite{ashby1970deformation}.\\
Modeling phenomena such as size effects is an ongoing effort and can not be achieved by using classic plasticity theories that lack an internal length scale. Gradient plasticity theories, however, (e.g.,~\cite{aifantis1984microstructural,gurtin2002gradient,gao1999mechanism}), were developed in the spirit of accounting for, e.g., the increase in the overall yield strength observed upon refining the grain size of the material or the specimen size. Therefore, an internal length scale is often introduced in a defect energy contribution to the free energy, thereby considering the gradients of, e.g., plastic slips. For the modeling of the interface resistance of GBs against dislocation movement, the two ideal limit cases of microhard and microfree conditions~\cite{gurtin2002gradient} are commonly considered. The microhard case corresponds to dislocations being restricted to remain within individual grains. In contrast, the microfree case corresponds to dislocations crossing the GBs unrestrictedly. The GB resistance in gradient plasticity theories can also be accounted for by using an additional interface (potential) term for the GBs in the free energy~\cite{aifantis2006interfaces}. For example, the plastic strain at the interface can be controlled with the additional potential (see also~\cite{fredriksson2005size}). Thereby, the yielding of GBs in the model can be decoupled from the yielding of the bulk material. The mechanism of GB yielding, thus, allows one to model GB behavior in between the two idealized microhard and microfree limits~\cite{wulfinghoff2013gradient,van2013grain}. Using GB yielding, however, it is not possible to distinguish between dislocation mechanisms on the continuum scale such as dislocations passing through the GB and dislocations depositing at the GB but originating from different grains~\cite{zhang2014}.\\
The gradient plasticity theory of~\cite{aifantis1999strain} models the experimentally observed size effect on thin copper wires from~\cite{fleck1994strain}. The authors fit their model to the individual experimental curves for each wire diameter. As a result of this procedure, the possible variations in texture are accounted for via the introduced gradient coefficients. Thus, different sets of parameters need to be used for each experimental data curve.\\
The experimental results by~\cite{fleck1994strain} are also modeled to some extent by the gradient plasticity approach~\cite{al2006physically}. A linear coupling between SSD- and GND-densities in the Taylor-type relation for the flow stress is assumed, there. However, this approach gives less agreement with the experimental data than the harmonic sum utilized for the coupling in~\cite{fleck1994strain} (see also the references on this in~\cite{al2006physically}).\\
Size effects can also be modeled using the critical thickness theory (CTT), see~\cite{dunstan2009elastic}. There, it is found that the consideration of only the specimen and the grain size in the model gives good agreement with the presented experimental data of twisted thin copper wires. However, the model was validated against data of microwires with only two different diameters and did not include the texture.\\
A phenomenological gradient plasticity theory~\cite{fleck2009mathematicala,fleck2009mathematicalb} is used in~\cite{idiart2010size} to analyze the torsion of twisted elastic-plastic thin wires. It is argued, there, that strain gradients are the consequence of the interaction of various mechanisms. These include, for example, the resistance of GBs against plastic flow and the imposed strain field resulting from the torsion loading.\\
The cyclic torsion tests on thin polycrystalline gold and copper wires by~\cite{Liu2015} show that energetic and dissipative contributions are necessary for the gradient plasticity modeling of certain phenomena. To account for this, the authors extend the framework~\cite{fleck2009mathematicala,fleck2009mathematicalb} by a gradient-related (and thus GND-related) energy of cold work.\\
In an effort to isolate the imposed strain gradient by torsion loading, the misorientation distribution within the single-crystalline cross sections of thin bamboo-structured gold wires is experimentally investigated for torsion loading in~\cite{ziemann2015deformation}. In addition, the equivalent plastic strain distribution within the cross sections is determined by crystal plasticity simulations. The results of both approaches show the overall strain gradient imposed by the torsion loading \mw{without disturbing contributions of individual grain boundaries}.\\
Single-crystalline copper is investigated in~\cite{kaluza2011torsion}. There, an analytical solution of a continuum theory of dislocations is obtained for torsion of a rod with circular cross section. The analytical results show good agreement with the experimental data presented up to moderate twists.\\
The torsion and tensile experiments on polycrystalline copper microwires by~\cite{liu2012size} \mw{are in line with} former results by~\cite{fleck1994strain}. \mw{The interpretation given by the authors was that the observed substantial size effect under torsion loading (but minor size effect under tensile loading) should mainly be associated to the GNDs.} \mw{Microwires with similar average grain sizes were used in their study.} \mww{However, the influence of a specific texture of the microwires was not considered.}\\
Observations in regard to the different size effects on torsion and tensile loaded thin wires are also made in the work of~\cite{liu2013toward}. There, the phenomenological gradient plasticity theories of Fleck and Hutchinson (FH)~\cite{fleck1997strain,fleck2001reformulation}, and Chen and Wang (CW)~\cite{chen2000new} as well as Aifantis~\cite{aifantis1999strain} (see also the references in~\cite{liu2013toward}) are compared. The ability of the theories to predict the size effects present in the experiments of~\cite{fleck1994strain} and present in the own experimental data of the authors is tested. These theories are contrasted with the CTT used in~\cite{dunstan2009elastic} (see also the references in~\cite{liu2013toward}). It is concluded in~\cite{liu2013toward}, that the FH~\cite{fleck1997strain,fleck2001reformulation} and the CW~\cite{chen2000new} theories both are more in-line with the experiment. Both theories account for the size effect of an increase in initial yield strength and exhibit a smaller influence on the strain hardening, while the Aifantis model~\cite{aifantis1999strain} opposes these trends. The initial yield size effect is mainly caused by the external geometrical size constraints put on the volume. As outlined in~\cite{liu2013toward}, this behavior can be explained with the CTT~\cite{dunstan2009elastic}. Furthermore, it is concluded that the internal length scale is not a constant (a finding that has been confirmed in other works in the past, too, see~\cite{al2006physically} and references therein for a more detailed discussion).\\
The model of~\cite{rahaeifard2014strain} is able to account for the yield strength increase in both the experimental data~\cite{fleck1994strain} and~\cite{liu2013toward} by using a size dependent yield criterion.\\
In~\cite{bardella2015modelling}, the gradient plasticity theory of~\cite{gurtin2005theory} is implemented and compared to experimental results~\cite{fleck1994strain}, exhibiting good agreement with the size effect for some of the investigated thin wires. Good agreement with the analytical solution by~\cite{chiricotto2012torsion} is found, too.\\
In the recent experimental work \mw{by Chen et al.~\cite{chen2015size}, the deformation behavior of coarse grained Au microwires with different diameters was} investigated for both torsion and tensile loading. A pronounced size effect is found for the torsion loading case, but an inverse size effect is observed for the tensile loading case. It \mw{is concluded} that these contrary behaviors observed are due to an enhancement of the size effect by specific microstructural characteristics. Therefore, these size effects are investigated in the present contribution using a gradient plasticity framework~\cite{wulfinghoff2013gradient}. Contrary to most approaches undertaken in the literature, the influence of both the texture and the grain size is considered.\\
A shortcoming of many gradient plasticity models are the arising high computational costs of 3D-simulations, a consequence of the consideration of, e.g., the gradients of all plastic slips in the model. The theory of~\cite{wulfinghoff2012equivalent,wulfinghoff2013gradient}, however, circumvents this issue by using an equivalent plastic strain as additional degree of freedom, instead of all plastic slips. Within a micromorphic implementation (see~\cite{forest2009micromorphic}, and~\cite{bayerschen2015equivalent} for a brief overview on micromorphic modeling) the computational costs of the model are reduced such that fully three-dimensional simulations of grain aggregates can be performed within reasonable computational times. Therefore, this model can be used to carry out the extensive simulations necessary to investigate the experimentally observed size effects under different loading conditions by~\cite{chen2015size}. The employed model has been shown in~\cite{wulfinghoff2013gradient} to reproduce the grain size effects by~\cite{yang2012yield} using oligocrystalline grain aggregates with similar average grain sizes as in the experiments. The texture influence, however, has not been investigated in detail, see~\cite{wulfinghoff2013gradient}. \rtwoC{In this model, a Voce-type hardening relation for bulk materials (see, e.g., \cite{voce1948relationship,voce1955practical,kocks1976laws}) is combined with a grain boundary yield criterion.} With this approach, the occurring size effects are mainly influenced by the grain boundary yield strength. Recently, the GB yield condition has been extended to account for GB hardening~\cite{bayerschen2015equivalent}. In the present contribution, however, hardening of the GBs is neglected due to a lack of data such as plastic strain profiles.
\topic{Outline} At first, the experimental results under consideration,~\cite{chen2015size}, are summarized and the texture distribution in several cross sections of the microwires is evaluated. Then, an overview on the utilized gradient plasticity model is given. The finite element implementation, boundary conditions and the discretization of the specimens are briefly addressed. Model parameters are calibrated to the experimental tensile test results and with these parameters the torsion tests are simulated, subsequently. The influence of a variation in the specimen texture is investigated by calibrating the model parameters for a texture with equal area shares of both orientations considered, at first, and for the simplified evaluated texture distributions from the experiments, afterwards. Discussion of the results is followed by the conclusions.
\topic{Notation} A direct tensor notation is preferred throughout the text. Vectors and 2nd-order tensors are denoted by bold letters, e.g., by \f{\fa} or~\f{\fA}. A linear mapping of 2nd-order tensors by a 4th-order tensor is written as \f{\fA=\ffC[\fB]}. The scalar product and the dyadic product are denoted, e.g., by~\f{\fA\cdot\fB} and~\f{\fA\otimes\fB}, respectively. The composition of two 2nd-order tensors is formulated with~\f{\fA\fB}. The 2nd-order unity tensor is denoted by~\f{\fI}. Matrices are denoted by a hat, e.g., by \f{\hat\varepsilon}.
\section{Experiments}\label{sec:exp}
\begin{figure*}[htbp]
\centering
 \includegraphics[width=0.95\linewidth]{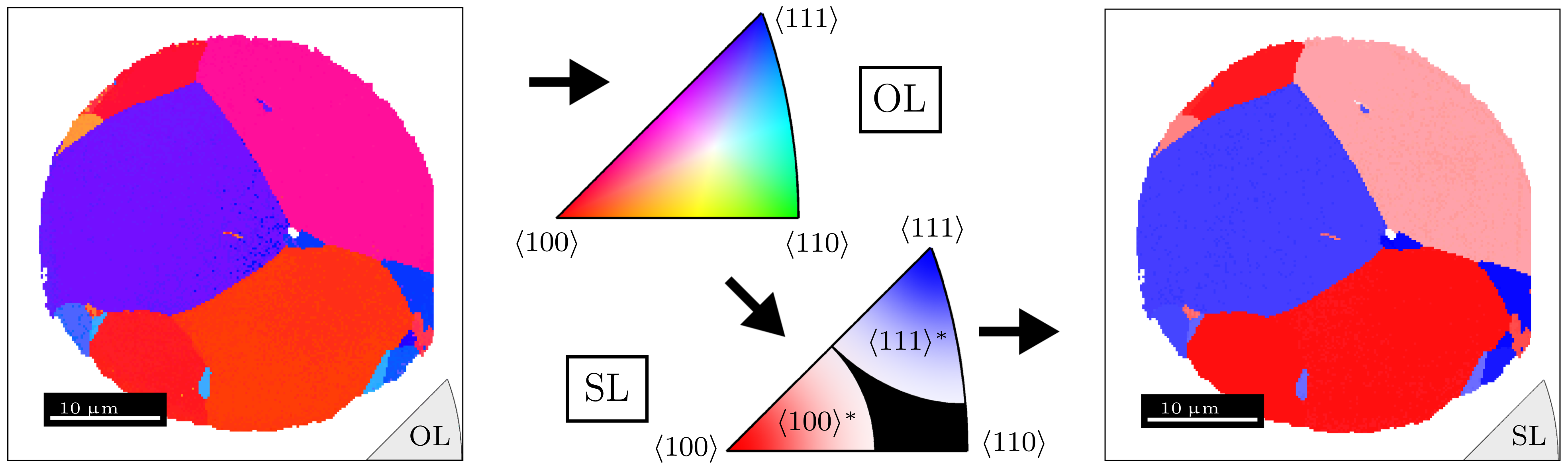}\\[1.5ex]
 \begin{minipage}{.32\textwidth}
  \hspace{23mm}
{\bf (a)}
 \end{minipage}
 \begin{minipage}{.32\textwidth}
  \hspace{23mm}
{\bf (b)}
\end{minipage}
 \begin{minipage}{.32\textwidth}
  \hspace{23mm}
{\bf (c)}
\end{minipage}\\[2ex]
 \includegraphics[width=0.95\linewidth]{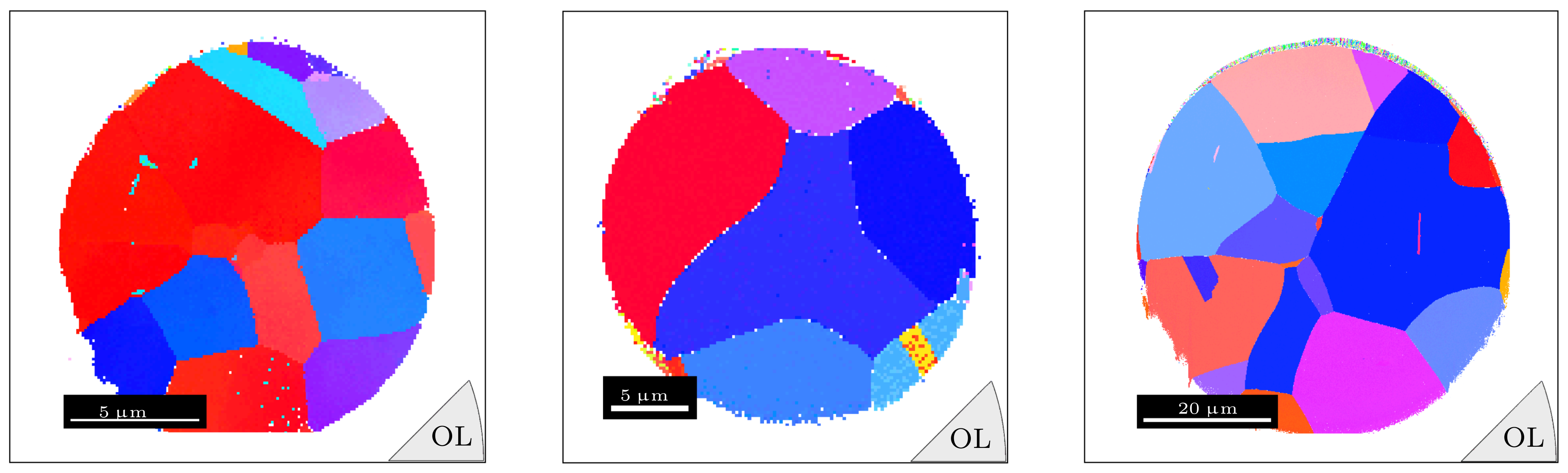}\\[1.5ex]
 \begin{minipage}{.32\textwidth}
  \hspace{23mm}
{\bf (d)}
 \end{minipage}
 \begin{minipage}{.32\textwidth}
  \hspace{23mm}
{\bf (e)}
\end{minipage}
 \begin{minipage}{.32\textwidth}
  \hspace{23mm}
{\bf (f)}
\end{minipage}
\caption{Top row: approach to simplify textural features using the example of an Au-wire with a diameter of \f{40\,\upmu}m: {\bf a} original EBSD map, {\bf b} data points revealing a misorientation of \f{\leq27.37^\circ} related to either the \f{\langle100\rangle}-axis or \f{\langle111\rangle}-axis, respectively, are defined as \f{\langle100\rangle^*}- or \f{\langle111\rangle^*}-oriented spots (data points located outside the defined areas are ignored),  {\bf c} simplified microstructure. Bottom row: original EBSD maps from wires with a diameter of {\bf d} \f{15\,\upmu}m, {\bf e} \f{25\,\upmu}m, and  {\bf f} \f{60\,\upmu}m. (OL: original legend, SL: simplified legend)}
\label{fig:exp_fig}
\end{figure*}

\mw{In order to study the impact of microstructural features on the mechanical behavior of oligocrystalline Au microwires of different thickness in tension and torsion, Chen and co-workers already performed several focused ion beam and electron backscattered diffraction (EBSD) scans on cross-sections of samples by using a Dual Beam facility from FEI (Nova NanoLab 200), equipped with an EBSD system from Oxford Instruments \cite{chen2015size}. \mww{Besides the determination of average grain sizes, orientation maps were used to identify textural differences. However, at that time} the maps were only assessed in a qualitative manner. Since both the average grain size and the textural features are considered in the present gradient plasticity studies, the missing quantitative examinations of the grain orientation distribution in dependence of the wire diameter were carried out within the frame of this work.}
Therefore, the original EBSD maps from the investigated wires (see \figref{fig:exp_fig}a and \figref{fig:exp_fig}d-f; generally up to four different cross sections for every wire diameter) were reanalyzed by using MATLAB R2014a with MTEX toolbox (see \cite{hielscher2008novel,bachmann2010texture}) in Version~4.1.1.
For the gradient plasticity simulations, the simplification is made that the wires consist only of ideally \f{\langle100\rangle}- and \f{\langle111\rangle}-oriented grains. Taking this into account, the experimental data were prepared such that the orientation of each spot within a cross section was counted either as \f{\langle100\rangle^*} or as \f{\langle111\rangle^*} normal to the surface when its misorientation is located within a radius of \f{27.37^\circ} related to either one of the both main directions (see \figref{fig:exp_fig}b-c). The value of \f{27.37^\circ} was chosen based on the fact that it is half the angle between the \f{\langle100\rangle}- and the \f{\langle111\rangle}-axis (points with orientations lying outside of the two circles are generally rare and, thus, neglected). Subsequently, the entire fractions of \f{\langle100\rangle^*}- as well as \f{\langle111\rangle^*}-oriented areas were determined by counting the number of spots for both orientations. Finally, the average area shares for every wire diameter were calculated. \mww{These results are shown in Table~\ref{tab:exps} together with the average grain sizes (determined \mw{by using a modified} line-interception method), both already published in \cite{chen2015size}. \rtwoC{It should be noted that only the average grain size is considered as microstructural parameter since the tensile flow stresses of the microwires exhibit comparable values \cite{chen2015size} in spite of differences in the grain sizes and the grain size distributions of the different wires.}}\\
\begin{table*}[htbp]
\renewcommand{\arraystretch}{1.2}
\caption{\mww{Microstructural features of coarse-grained Au microwires with different diameter ($^1$: values from \cite{chen2015size})}}
\label{tab:exps}
\centering
\begin{tabular}{ccccc}
\hline
\mww{Nominal wire diameter}&Real wire diameter$^1$&Avg.\ grain size$^1$&Area share&Area share\\
\mww{\f{D}} ($\upmu$m)&\mww{\f{D_{\rm exp}}} ($\upmu$m)&\f{d_{\rm avg, exp}} ($\upmu$m)&\f{\langle111\rangle^*} (\%) &\f{\langle100\rangle^*} (\%) \\\hline
15&14.78&2.96&33&67\\
25&24.46&3.49&45&55\\
40&37.38&8.63&67&33\\
60&59.53&6.38&75&25\\\hline
\end{tabular}
\end{table*}
\mww{When comparing the individual area shares of the simplified microstructures with the grain colour distributions within all related original EBSD maps, one may suppose that the calculated values are matching comparably well for the two wires of smaller diameter. In contrast, it seems that the simplified grain orientation distributions of the two wires with bigger diameter do not reflect the given textural features well enough. Here, larger fractions of grains with stronger misorientations related to the ideal \f{\langle100\rangle}-axis or the ideal \f{\langle111\rangle}-axis exist. The smaller wires, however, mainly consist of almost ideally \f{\langle100\rangle}- and \f{\langle111\rangle}-oriented grains (compare, for example, \figref{fig:exp_fig}a to \figref{fig:exp_fig}f). \rtwoC{This finding was also confirmed by results of Young's modulus investigations by \cite{chen2015size}.} As a consequence, additional gradient plasticity simulations are carried out for the microstructures with equal fractions of \f{\langle100\rangle}- and \f{\langle111\rangle}-oriented grains. \revM{Thus, the influence of inadequately described microstructure on the size effect can be studied.}}
\section{Mathematical framework}
\newcommand{\rev}{\color{black}}
\newcommand{\rone}{\color{black}}
\newcommand{\roneC}[1]{{\color{black}{#1}}}
\newcommand{\caq}{{\roneC{\zeta}}}
\newcommand{\dcaq}{{\roneC{\dot\zeta}}}
\newcommand{\redsw}{\color{black}{}}
\subsection{Basic assumptions}
The gradient plasticity framework used in the work at hand is described in detail in~\cite{wulfinghoff2013gradient}. Here, only a brief summary is given. A small strain framework is used, and the strain~\f{\feps} is assumed to be the symmetric part of the gradient of the displacement~\f{\fu}. The additive decomposition of~\f{\feps} into an elastic contribution~\f{\feps^{\rm e}} and a plastic contribution~\f{\feps^{\rm p}} is presumed. The plastic strain~\f{\feps^{\rm p}} is given by the sum over the slip parameters~\f{ \lambda_\alpha } multiplied by their respective Schmid tensors~\f{\fM_\alpha^s}, where the slip plane normals are denoted by~\f{\fn_\alpha}, and the slip directions are~\f{\fd_\alpha}. A face-centered cubic crystal structure is considered, and due to the non-decreasing slip parameters (\f{\dot\lambda_\alpha\geq0}) used, the identifier~\f{\alpha} of the slip systems ranges from~\f{\alpha=1,\ldots,24}. Summation of the slip parameters~\f{ \lambda_\alpha } gives the equivalent plastic strain~\f{ \gaq(\hat\lambda) }, cf.\ Table~\ref{tab:BasicAssumptions}. The micromorphic counterpart of the equivalent plastic strain is denoted by~\f{\zeta}, in the following.
\begin{table}[htbp]
\renewcommand{\arraystretch}{1.4}
\caption{Summary of the basic assumptions of the gradient plasticity model}
\label{tab:BasicAssumptions}
\centering
\begin{tabular}{l|rl}
\hline
Displacement gradient                                            & \f{\nabla\fu}           & = \f{\partial_{x_j}u_i\fe_i\otimes\fe_j}\\
Strain tensor                                                    & \f{\feps}               & = \f{\fsym{\nabla\fu}}\\
Additive decomposition                    & \feps                   & = \f{\feps^{\rm e} + \feps^{\rm p}}\\
Plastic strain tensor                        & \f{ \feps^{\rm p} }     & = \f{\sum_\alpha\lambda_\alpha\fM_\alpha^s },\\ 
Symmetric Schmid tensor		& \f{\fM_\alpha^s}        &= \f{ \fsym{\fd_\alpha\otimes\fn_\alpha}} \\
Equivalent plastic strain                                        & \f{ \gaq(\hat\lambda) } & = \f{ \sum_{\alpha} \lambda_\alpha },\\ 
Slip parameters                                                 &\f{ \lambda_\alpha }    &= \f{ \int  \dot \lambda_\alpha  \d t, \dot \lambda_\alpha \geq 0}  \\
\hline
\end{tabular}
\end{table}
\subsection{Field equations and boundary conditions.}
Based on the principle of virtual power, e.g.,~\cite{Maugin1980}, the field equations and boundary conditions, summarized in Table~\ref{tab:FieldEquationsAndBCs}, can be derived according to a standard procedure (e.g., ~\cite{Gurtin2007}). The balance of linear momentum is complemented by an additional microforce balance and the GB microtraction balance. In this context, the jump of the gradient stress \f{\fxi} is written as \f{\llbracket\fxi\rrbracket=\fxi^+-\fxi^-}. Here, the normal~\f{\fn_\Gamma} of the GB points from ``-'' to ``+''. In analogy to the Neumann BCs for the Cauchy stress, BCs for the gradient stress need to be prescribed, too, see Table~\ref{tab:FieldEquationsAndBCs}. It is remarked that \f{\pi} is a scalar and \f{\fxi} is a vectorial higher-order stress. They are work conjugate to the rates \f{\dot\zeta}, and \f{\nabla\dot\zeta}, respectively. The microtraction prescribed for the vectorial higher-order stress on the outer boundary is denoted by \f{\bar\Xi}, and the microtraction for the GBs is \f{\Xi_\Gamma}.
\begin{table}[htbp]
\renewcommand{\arraystretch}{1.2}
\caption{Field equations and boundary conditions of the gradient plasticity model}
\label{tab:FieldEquationsAndBCs}
\centering
\begin{tabular}{l|rl}
\hline
Linear momentum balance                     & \f{\fzero}          & \f{=\div{\fsigma}}, \f{\forall\fx\in\cB}\\
Microforce balance                          & \f{{\rev \pi}}      &\f{=\div{\fxi}},  \f{\forall\fx\in\cB\setminus\Gamma}\\
GB microtraction balance                           & \f{\Xi_\Gamma}      &\f{=\llbracket\fxi\rrbracket\cdot\fn_\Gamma}, \f{\forall\fx\in\Gamma}\\
Neumann BCs for:&&\\
\quad Cauchy stress              & \f{\fsigma \fn}     & \f{= \bar\ft},  \f{\mathrm{on}\ \partial \cB_t}\\
\quad Gradient stress  &\f{\fxi \cdot \fn}   &\f{= \bar \Xi}, \f{\mathrm{on} \ \partial \cB_{\,\Xi}}
\\\hline
\end{tabular}
\end{table}
\subsection{Free energy}
In the following, an additive decomposition of the free energy~\f{W} of the body~\f{\cB} into a volumetric contribution \f{W_{\rm V}} from the bulk and a contribution \f{W_\Gamma} per unit area of the GBs is assumed. The free energy reads
\begin{equation}
\begin{split}
  W = \int \limits_\cB W_{\rm V} \d v + \int \limits_{\Gamma} W_\Gamma \d a = \int \limits_\cB \left(W_{\rm e}(\feps,\feps^{\rm p}(\hat\lambda))+W_{\rm h}(\caq)\right.\\
\left.+W_{\rm g}(\nabla\caq)+\roneC{W_{\chi}(\zeta-\gamma_ {\rm eq}(\hat\lambda))}\right) \d v + \int \limits_{\Gamma} W_\Gamma(\caq) \d a,
\end{split}
\end{equation}
where~\f{\Gamma} is the union of all GBs. The volumetric contribution consists of the elastic energy density~\f{W_{\rm e}}, an isotropic hardening contribution~\f{W_{\rm h}}, a defect energy density~\f{W_{\rm g}} and the energy density~\f{W_{\chi}} coupling the micromorphic variable~\f{\zeta} to the field variable~\f{\gamma_ {\rm eq}} (see~\cite{wulfinghoff2013gradient} for details). The energy densities are given in detail in Table~\ref{tab:FreeEnergyDensities}. Grain boundary influence in the material is accounted for by the area-specific energy density~\f{W_\Gamma}. In combination with a yield criterion for the GBs, dislocation transfer behavior across GBs in between the idealized limits of free dislocation movement across the GBs and fully impeded movement can be modeled.\\
The hardening associated to SSDs is modeled by the isotropic energy density~\f{W_{\rm h}} of the bulk, and the influence of the GNDs is accounted for by~\f{W_{\rm g}} in combination with the GB yield condition. Due to the energetic modeling approach taken, the gradient stress is obtained by~\f{\fxi=\partial W_{\rm g}(\nabla\zeta)/\partial\nabla\zeta}. It is remarked that the consideration of the hardening contribution in the free energy is a purely convenient choice, allowing for an energetical stress-measures derivation. Instead, the hardening could be considered by postulating dissipative stress terms without contributions to the free energy.
\begin{table*}[htbp]
\renewcommand{\arraystretch}{1.3}
\caption{Free energy densities of the gradient plasticity model}
\label{tab:FreeEnergyDensities}
\centering
\begin{tabular}{l|l|c}
\hline
\multicolumn{2}{c|}{ Energy densities }                                                                                                                                                  &  Type \\\hline
\multirow{1}{*}{ \f{W_{\rm e}(\feps,\feps^{\rm p}(\hat\lambda))} }                             & \f{ (\feps-\feps^{\rm p}(\hat\lambda))\cdot\ffC[\feps-\feps^{\rm p}(\hat\lambda)]/2 }    & Elastic \\
\multirow{1}{*}{ \f{W_{\rm h}(\caq)}         }                                                 & \f{ (\tau_\infty^{\rm C}-\tau_0^{\rm C})\caq+(\tau_\infty^{\rm C}-\tau_0^{\rm C})^2\exp\left(-\Theta\caq/\left(\tau_\infty^{\rm C}-\tau_0^{\rm C}\right)\right)/\Theta } & Isotropic hardening \\
\multirow{1}{*}{ \f{W_{\rm g}(\nabla\caq)} }                                                   & \f{ K_{\rm G}\nabla\caq\cdot\nabla\caq /2}                                      & Defect \\
\multirow{1}{*}{ \f{W_\chi(\zeta-\gamma_ {\rm eq}(\hat\lambda)) } }                            & \f{ H_\chi(\zeta-\gamma_{\rm eq})^2/2}                                                   & Numerical coupling\\
\multirow{1}{*}{ \f{W_\Gamma(\caq)}}  & \f{\Xi_0^{\rm C}\caq }                                      & Grain boundary\\\hline
\end{tabular}
\end{table*}\\
The elastic stiffness tensor is denoted by~\f{\ffC}, the initial yield stress (of a slip system) is~\f{\tau_0^{\rm C}}, the respective saturation stress is~\f{\tau_\infty^{\rm C}}, and the initial hardening modulus is~\f{\Theta}. An internal length scale is introduced in the model by the defect parameter~\f{K_{\rm G}}. The constant GB yield strength is denoted by~\f{\Xi_0^{\rm C}}. For the coupling of the micromorphic variable to the equivalent plastic strain the penalty parameter~\f{H_\chi} is used.
\subsection{Dissipation and flow rules}\label{sec:diss}
After applying the equivalence of internal and external power and neglecting thermal effects, the total dissipation can be written as
\begin{equation}
 \cD_{\rm tot} = \int \limits_\cB \cD \d v + \int \limits_{\Gamma} \cD_\Gamma \d a \geq 0 .
\end{equation}
In accordance with~\cite{Cermelli2002}, the dissipation of the bulk is assumed to be caused by the dissipative shear stresses of the slip systems, i.e., \f{\cD=\sum_\alpha\tau_\alpha^{\rm d}\dot\lambda_\alpha}. Using the basic assumptions from Table~\ref{tab:FieldEquationsAndBCs}, employing the difference between the external power and the rate of the stored energy, using the microforce balance, and the resolved shear stresses~\f{\tau_\alpha = \fsigma \cdot \fM^{\rm S}_\alpha}, the dissipative shear stresses are given by 
\begin{equation}
 \tau_\alpha^{\rm d}=\tau_\alpha+\div\fxi-\beta.
\end{equation}
Here, \f{\beta=\partial W_{\rm h}(\zeta)/\partial\zeta} denotes the isotropic hardening stress. The GB dissipation per unit area is 
\begin{equation}
  \cD_\Gamma = \Xi_\Gamma^{\rm d}\, \dcaq = (\Xi_\Gamma - \Xi_\Gamma^{\rm e}) \, \dcaq \geq 0,
\end{equation}
with the energetic microtraction~\f{\Xi_\Gamma^{\rm e} = \partial_{\caq} W_\Gamma} and the dissipative microtraction~\f{\Xi_\Gamma^{\rm d}}. An overstress-type flow rule is assumed for the bulk,
\begin{equation}
 \label{eq:powerLaw}
 \dot \lambda_\alpha = \dot \gamma_0 \left\langle
\frac{\tau_\alpha^{\rm d} - \tau^{\rm C}_0}{\tau^{\rm D}}
\right\rangle^p
=\dot \gamma_0 \left\langle
\frac{\tau_\alpha+\div\fxi- ( \tau^{\rm C}_0 + \beta )}{\tau^{\rm D}}
\right\rangle^p.
\end{equation}
Here,~\f{\dot \gamma_0} denotes the reference shear rate,~\f{\tau^{\rm D}} the drag stress and~\f{p} the sensitivity exponent. The yield function~\f{f_\Gamma} of the GBs is assumed to be
\begin{equation}
  f_\Gamma = \Xi_\Gamma^{\rm d} = [\![ \fxi ]\!] \cdot \fn -\Xi_\Gamma^{\rm e}.
\label{eq:gb_yield_function}
\end{equation}
The corresponding Kuhn-Tucker conditions are ${f_\Gamma\leq0}$, $\dcaq\geq0$, ${\dcaq f_\Gamma=0}$, and the GB yield strength is obtained from~\f{\Xi_\Gamma^{\rm e}= \partial_{\caq} W_\Gamma}.
\section{Finite element simulations}
\subsection{Implementation}
\subsubsection{Simulation specimens and discretization}
Finite element (FE) simulations are performed on simplified grain aggregates with diameter~\f{D} in the range of~\f{15-60\,\upmu}m, according to the experiments. The length~\f{l} of the simulation volume for each diameter is chosen such that~\f{l/D=1.25=const.}, in order to allow for a comparison of the results obtained for different diameters (see Table~\ref{tab:disc} for the diameters~\f{D}).
\begin{table}[htbp]
\renewcommand{\arraystretch}{1.2}
\caption{Discretizations of oligocrystalline gold microwires. Abbreviations used are: DOFs - degrees of freedom, ECD - equivalent circle diameter, FEs - finite elements}
\label{tab:disc}
\centering
\begin{tabular}{cccc}
\hline
Diameter&{Average grain size}&FEs&DOFs\\
\f{D_{\rm sim}}&ECD: \f{d_{\rm avg,sim}}&&\\
($\upmu$m)&($\upmu$m)&&\\\hline
15&3.67&54720&235396\\
25&6.12&54864&236180\\
40&9.80&54720&235396\\
60&14.69&54864&236180\\\hline
\end{tabular}
\end{table}
The FE implementation of the theory at hand is incorporated in the micromorphic implementation outlined in~\cite{wulfinghoff2013gradient}. This implementation has been carried out in an in-house FE code of the Institute of Engineering Mechanics at KIT. An enhanced time-integration algorithm~\cite{wulfinghoff2013equivalent} is used that allows for significantly larger time steps than conventional algorithms. Exemplarily, the grain aggregate with smallest diameter has been investigated in a convergence study under tensile as well as under torsion loading. The discretizations of the simplified grain aggregates are chosen such that approximately doubling the degrees of freedom of the FE mesh leads to a reduction of less than 1.5\% in the relative error in the overall mechanical responses after the final time step of the grain aggregates. All discretizations are generated using ABAQUS-CAE as preprocessor and are subsequently used in the in-house FE code carrying out the numerical simulations. The computational times remain within the order of several hours for the employed discretizations, listed in \tabreff{tab:disc}. Due to the discretization algorithm, slightly different numbers of finite elements are used in the individual discretizations. However, no substantial influence on the overall mechanical response is caused by the resulting small differences in the degrees of freedom of the grain aggregates.
\subsubsection{Microstructure and texture used in the simulations}
\begin{figure*}[hbtp]
\begin{center}
 \includegraphics[width=0.95\linewidth]{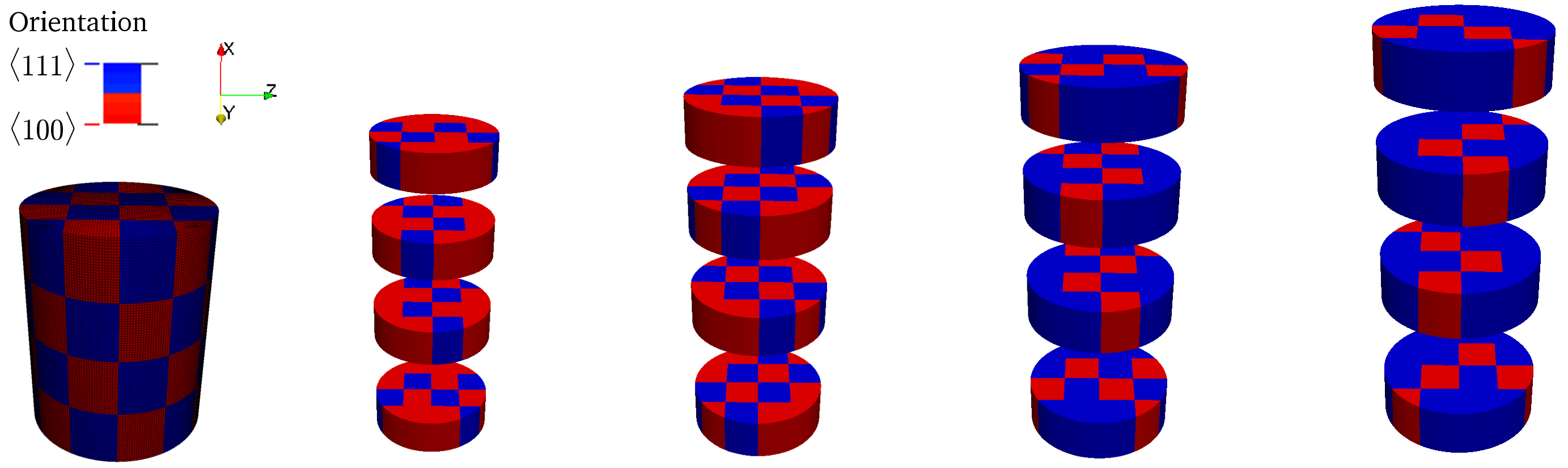}\\[1.5ex]
\hspace{0.08\textwidth} {\bf (a)} \hfill {\bf (b)} \hfill {\bf (c)} \hfill \quad {\bf (d)} \hfill \quad {\bf (e)}\quad\quad\quad\quad\hspace{0.08\textwidth}
\end{center}
\caption{Textures of oligocrystalline grain aggregates used in finite element simulations: {\bf a} equally distributed (alternating) \f{\langle100\rangle}- and \f{\langle111\rangle}-orientations, depicted are the Gauss point volumes of the employed finite element mesh, {\bf b} $15\,\upmu$m-aggregate with 20 \f{\langle111\rangle}-grains and 44 \f{\langle100\rangle}-grains, {\bf c} $25\,\upmu$m-aggregate with 24 \f{\langle111\rangle}-grains and 40 \f{\langle100\rangle}-grains, {\bf d} $40\,\upmu$m-aggregate with 44 \f{\langle111\rangle}-grains and 20 \f{\langle100\rangle}-grains, and {\bf e} $60\,\upmu$m-aggregate with 48 \f{\langle111\rangle}-grains and 16 \f{\langle100\rangle}-grains. The size of all grain aggregate visualizations has been rescaled \mw{for visualization. The specific number of \f{\langle100\rangle}- and \f{\langle111\rangle}-oriented grains for {\bf b} to {\bf e} are adapted to the experimental data in Table~\ref{tab:exps}}}
\label{fig:grains}
\end{figure*}

For the microstructure of the thin wires, grains of simplified shape are considered (see \figref{fig:grains}a).
The GBs are discretized by partitioning the simulation volume such that the GBs are planar surfaces composed of element surfaces in the finite element mesh. This is done by equidistant positioning of three grain boundary planes in each direction. Motivated by the observed average grain sizes of the oligocrystalline microwire specimens (see Table~\ref{tab:exps}), 16~grains are modeled within each of the four grain layers, i.e., 64 grains in the whole simulation volume. The resulting average grain sizes in terms of the equivalent circle diameter (ECD) are listed in \tabreff{tab:disc}.\\
Each grain is assigned either a \f{\langle100\rangle}- or a \f{\langle111\rangle}-orientation with regard to the specimen coordinate system (see also \figref{fig:grains}). This means that for \f{\langle100\rangle}-grains, the \f{\langle100\rangle}-crystal directions are aligned with the Cartesian coordinate system, while for \f{\langle111\rangle}-grains the \f{\langle111\rangle}-crystal direction is aligned with the central specimen axis. At first, it is assumed that equal area shares of both orientations are present in the cross sections and the orientations are assigned to the grains such that \f{\langle100\rangle}- and \f{\langle111\rangle}-orientations are alternating (see \figref{fig:grains}a).\\
However, in a re-evaluation of the orientation distribution of several cross sections from the experiments, a trend towards more \f{\langle100\rangle}-grains in the cross sections investigated has been observed for smaller wire diameters. Therefore, the texture of the microwires is accounted for, subsequently. For each wire diameter, the crystal orientations of several grains within each cross section are changed, e.g., from \f{\langle111\rangle}- to \f{\langle100\rangle}-orientation, until the area shares of the \f{\langle100\rangle}- and the \f{\langle111\rangle}-orientation are approximately equal to the corresponding experimentally determined simplified area shares (\figref{fig:grains}b-e).\\
As it can be seen, e.g., in \figref{fig:exp_fig}d, grains of very similar but not necessarily identical orientations are present in the cross sections of the microspecimen. Therefore, the interfaces between, e.g., two \f{\langle111\rangle}-grains of the texture are treated as GBs, nevertheless, in the simulations. The small variations in, e.g., the \f{\langle111\rangle}-orientation of such grains in the experiments are, however, not accounted for explicitly in the modeling approach. The strict classification into \f{\langle111\rangle}- and \f{\langle100\rangle}-orientations is justifiable for the two microwires with smaller diameters (see \secref{sec:exp}). For the two larger microwires, however, several crystal orientations in between these two orientations have been observed (see also \secref{sec:exp}). Therefore, two cases are investigated for each of the two microwires with bigger diameter: using equal area shares of both orientations (\figref{fig:grains}a), and using the simplified texture (\figref{fig:grains}d and \figref{fig:grains}e, respectively).
\subsubsection{Boundary conditions and simulation setup}
For the tensile loading, the bottom surface (in direction of the central specimen axis) is fixed in the loading direction, and on the top surface the displacement is prescribed until a nominal strain of \f{\varepsilon=0.05} is reached. Lateral contraction is allowed for by the boundary conditions in the transversal directions.\\
For the torsion loading, the top and bottom surface in direction of the central specimen axis are twisted in opposing directions until a \mww{maximum shear of \f{\gamma_{r=R}=0.02}} is reached. For this, Dirichlet conditions are imposed such that the displacements in the directions within the top and bottom planes (located at $x=const.$) are prescribed. However, the displacement in the direction of the central specimen axis is free, except at one finite element node (to ensure uniqueness of the position of the specimen in space).\\
In the beginning of the simulation, the equivalent plastic strain is set to vanish at all FE nodes. A grain boundary yield strength is then assigned to the respective nodes of the element surfaces coinciding with the grain boundary planes. On the simulation specimen surfaces, microfree conditions are imposed for the equivalent plastic strain, i.e., the equivalent plastic strain is not restricted, there. The determination of plastically active nodes on the GBs necessitates an additional algorithm, e.g., an active-set search (details in~\cite{wulfinghoff2013gradient}). In the beginning of the deformation, all nodes on the GBs are set to behave microhard, i.e., the GBs are plastically inactive, until the GB yield condition is fulfilled and they become plastically active.
\subsubsection{Material and model parameters}
Regarding the material parameters, literature values are used for the anisotropic elastic constants of gold (\f{C_{1111}=186}~GPa, \f{C_{1122}=157}~GPa, \f{C_{1212}=42}~GPa, \cip{roesler2006}). In all simulations, the employed reference shear rate is \f{\dot\gamma_0=10^{-3}/}s, the rate sensitivity exponent is~\f{p=20}, and the drag stress is~\f{\tau^{\rm D}=1}~MPa. The value of the penalty parameter, used to couple the micromorphic field variable~\f{\zeta} to the equivalent plastic strain~\f{\gaq}, is \f{H_\chi=10^8}~MPa. The length scale, resulting from a first estimate of the observed dislocation density of the microwires, is \f{l_{\rm int}=1/\sqrt{\rho_0}\approx3\,\upmu}m with \f{\rho_0\approx10^{11}}/m$^2$. The corresponding value for the defect energy parameter (\f{K_G=(l_{\rm int})^2E_{\rm avg}=0.6}~N, with the average Young's Modulus of all gold microwires \f{E_{\rm avg}\approx65}~GPa \mww{from \cite{chen2015size}}), however, leads to results substantially overestimating the experimental data in the elastic-plastic transition regime. Instead, the defect energy parameter~\f{K_G} is fitted to the experimental tensile curves by comparison of the slope in the elastic-plastic transition regime. This gives a value of \f{K_G=5\times10^{-4}}~N. The internal length scale resulting from this fit is \f{l_{\rm int,avg}=\sqrt{K_G/E_{\rm avg}}\approx0.09\,\upmu}m.\\
The remaining parameters are listed in the respective results sections. It is remarked that, at first, the same initial yield stress~\f{\tau_0^{\rm C}=1}~MPa (e.g.,~\cite{sachs1930zugversuche}) is used for all slip systems in the investigated cases. This value also can be motivated by using the Taylor-hardening relation \f{\tau_0^{\rm C}=\alpha b G\sqrt{\rho_0}} with \mw{average shear modulus \f{G_{\rm avg}\approx24}~GPa}, Burgers vector magnitude \f{b=0.224\times10^{-9}}m, and \f{\alpha=0.5}. Due to the additional yield criterion for the GBs, the overall yield strength in the model results from the combined behavior of both bulk and grain boundary yield strength. Later on, the initial yield stress for the slip systems and the GB yield strength are scaled for the case of torsion loading (details in \secref{sec:scaled}).\\
\revM{As it will be discussed in more detail, in the next sections, the GB yield strength of the model can not be used as a material constant when modeling the different microwires. This is due to the deviations of the grain sizes in the modeling from the experimental average grain sizes.}
\subsection{Numerical results}
\subsubsection{Gradient plasticity simulations using \mw{equal number of \f{\langle100\rangle}- and \f{\langle111\rangle}-oriented grains}}
\begin{figure*}[htbp]
\begin{center}
 \includegraphics[width=0.95\linewidth]{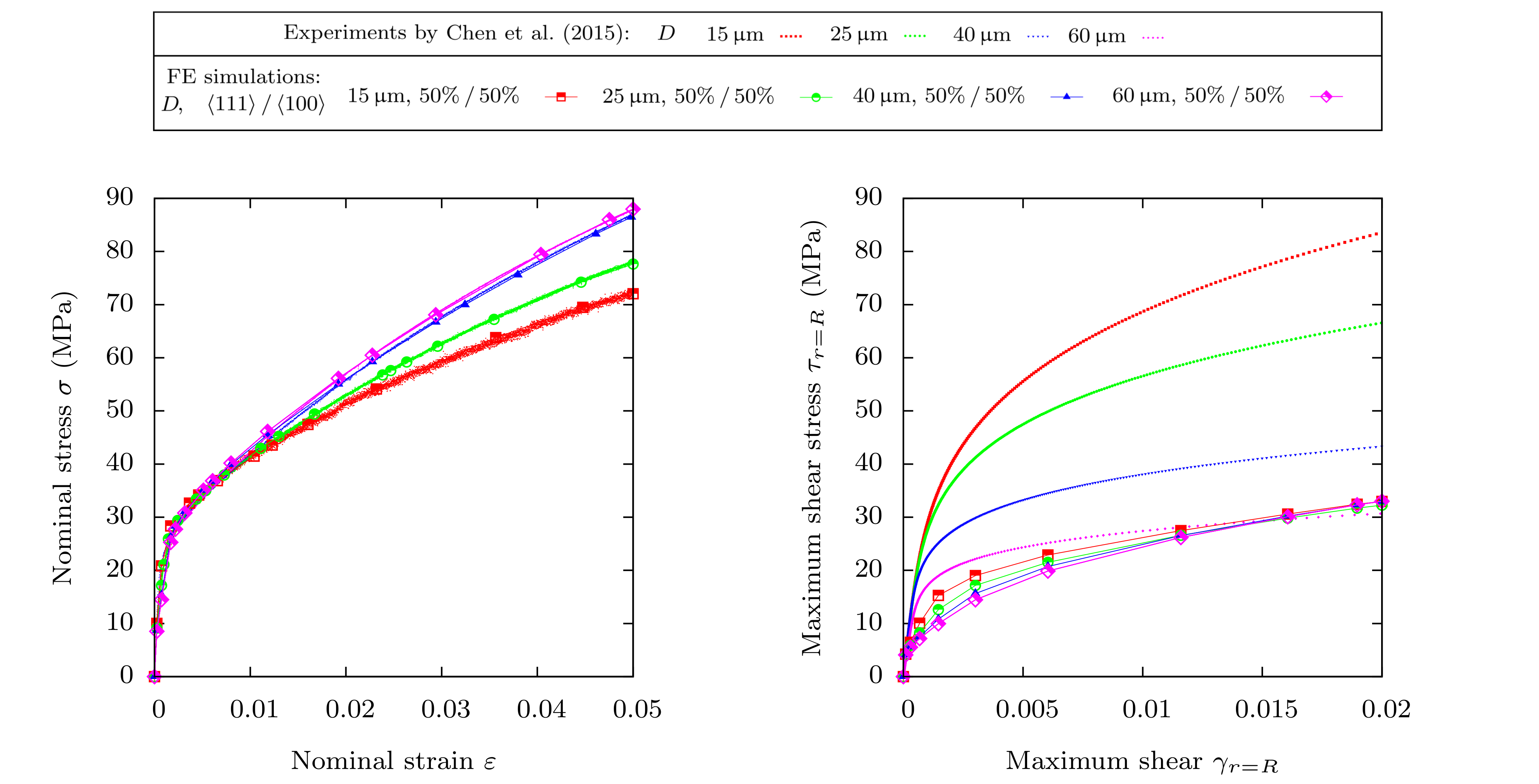}\\
 \begin{minipage}{.45\textwidth}
  \hspace{33mm}{\bf (a)}
 \end{minipage}
 \begin{minipage}{.45\textwidth}
  \hspace{40mm} {\bf (b)}
\end{minipage}
\end{center}
\caption{{\bf a} Stress-strain curves of tensile-loaded simplified grain aggregates in comparison to experimentally observed oligocrystalline microwire responses by~\cite{chen2015size}, model parameters calibrated to the tensile experiment. {\bf b} Maximum shear stress vs.\ maximum shear of the same aggregates under torsion loading compared to results by~\cite{chen2015size}, model parameters identical to {\bf a}. Equal cross-section area shares of \f{\langle111\rangle}- and \f{\langle100\rangle}-orientations used in the FE simulations}
\label{fig:size_50_50}
\end{figure*}

In the following, the simulations results of the grain aggregates modeling the gold microwires under tensile and torsion loading are presented. At first, equal distributions of \f{\langle111\rangle}- and \f{\langle100\rangle}-orientated grains are used, i.e., the diameter-dependent texture of the investigated specimens is not considered. The model parameters~\f{\tau_\infty^{\rm C}} (saturation stress of slip systems), \f{\Theta} (initial hardening modulus), and \f{\Xi_0^{\rm C}} (yield strength of the GBs) are fitted to the experimentally determined stress-strain curves (see \figref{fig:size_50_50}a for experimental data and tensile simulations with the fitted parameters).
Thereby, the set of parameters listed in Table~\ref{tab:zug_k_kG100_50_50_NEU} is obtained.
\begin{table}[htbp]
\centering
\renewcommand{\arraystretch}{1.2}
\caption{Model parameters calibrated to tensile data for identical cross-section area shares of \f{\langle111\rangle}- and \f{\langle100\rangle}-orientations}
\begin{tabular}{ccccc}
\hline
\f{D_{\rm sim}}&\f{\tau_0^{\rm C}}&\f{\Xi_{0}^{\rm C}}&\f{\Theta}&\f{\tau_\infty^{\rm C}}\\
($\upmu$m)&(MPa)&(Nm$^{-1}$)&(MPa)&(MPa)\\\hline
15&1&11&120&19\\
25&1&18&130&26\\
40&1&29&144&40\\
60&1&43&155&39\\
\hline
\end{tabular}
\label{tab:zug_k_kG100_50_50_NEU}
\end{table}
Using the same model parameters, the torsion response of the microwires is simulated, subsequently. The maximum shear stress \f{\tau_{r=R}=2M_T/(\pi R^3)} is plotted over the maximum shear~\f{\gamma_{r=R}} in \figref{fig:size_50_50}b. It is observable, there, that no pronounced size effect is obtained in the simulations of the torsion loading with equal area shares of both orientations and this set of parameters.
\subsubsection{Gradient plasticity simulations considering the simplified texture of the microwires}
\begin{figure*}[htbp]
\begin{center}
 \includegraphics[width=0.95\linewidth]{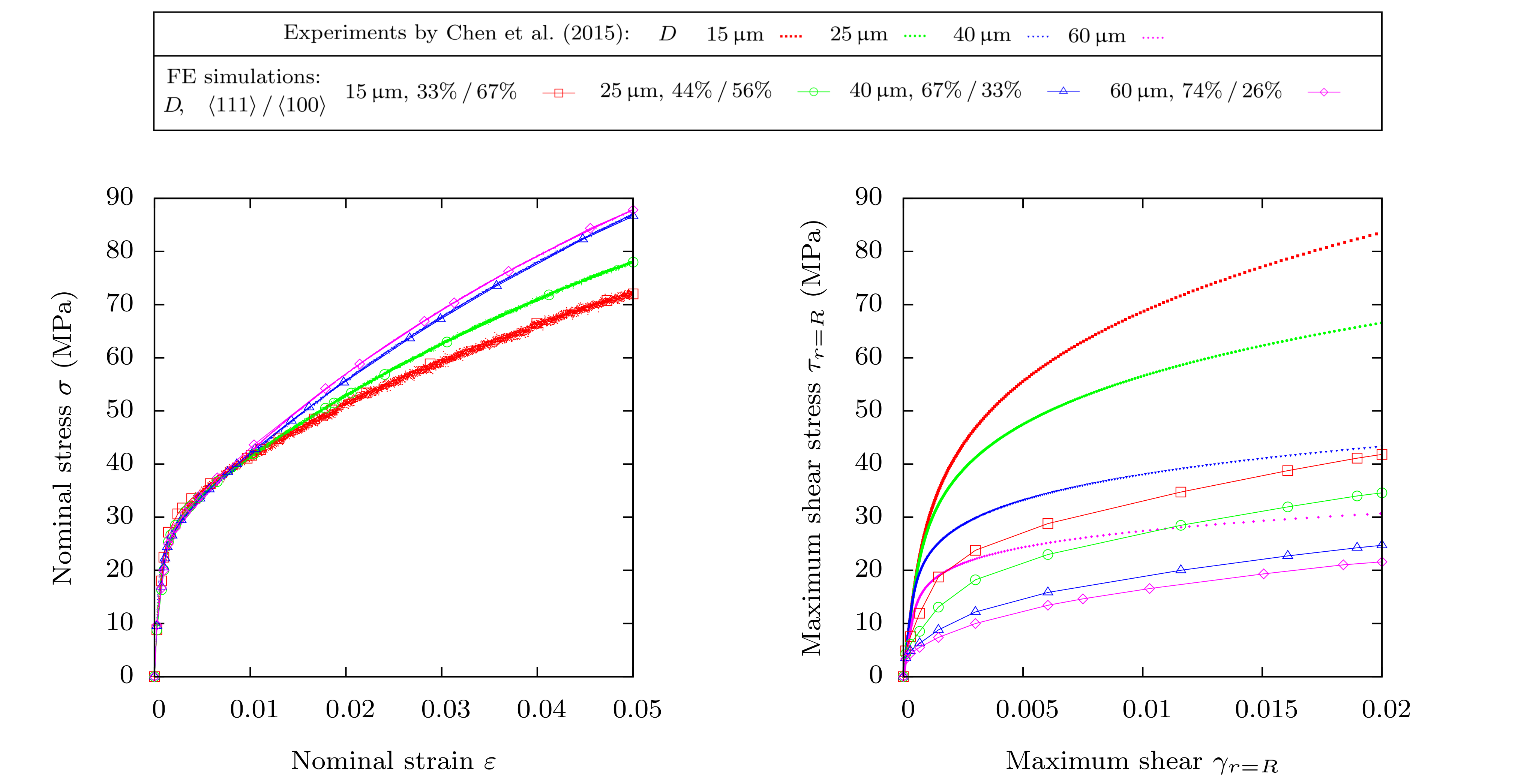}\\
 \begin{minipage}{.45\textwidth}
  \hspace{33mm}{\bf (a)}
 \end{minipage}
 \begin{minipage}{.45\textwidth}
  \hspace{40mm}{\bf (b)}
\end{minipage}
\end{center}
\caption{{\bf a} Stress-strain curves of tensile-loaded simplified grain aggregates in comparison to experimentally observed oligocrystalline microwire responses by~\cite{chen2015size}, model parameters calibrated to tensile experiment. {\bf b} Maximum shear stress vs.\ maximum shear of the same aggregates under torsion loading compared to results by~\cite{chen2015size}, model parameters identical to {\bf a}. The experimentally determined, simplified cross-section area shares of \f{\langle111\rangle}- and \f{\langle100\rangle}-orientations are used in the FE simulations}
\label{fig:size_70_30}
\end{figure*}

\begin{figure*}[htbp]
\begin{center}
 \includegraphics[width=0.95\linewidth]{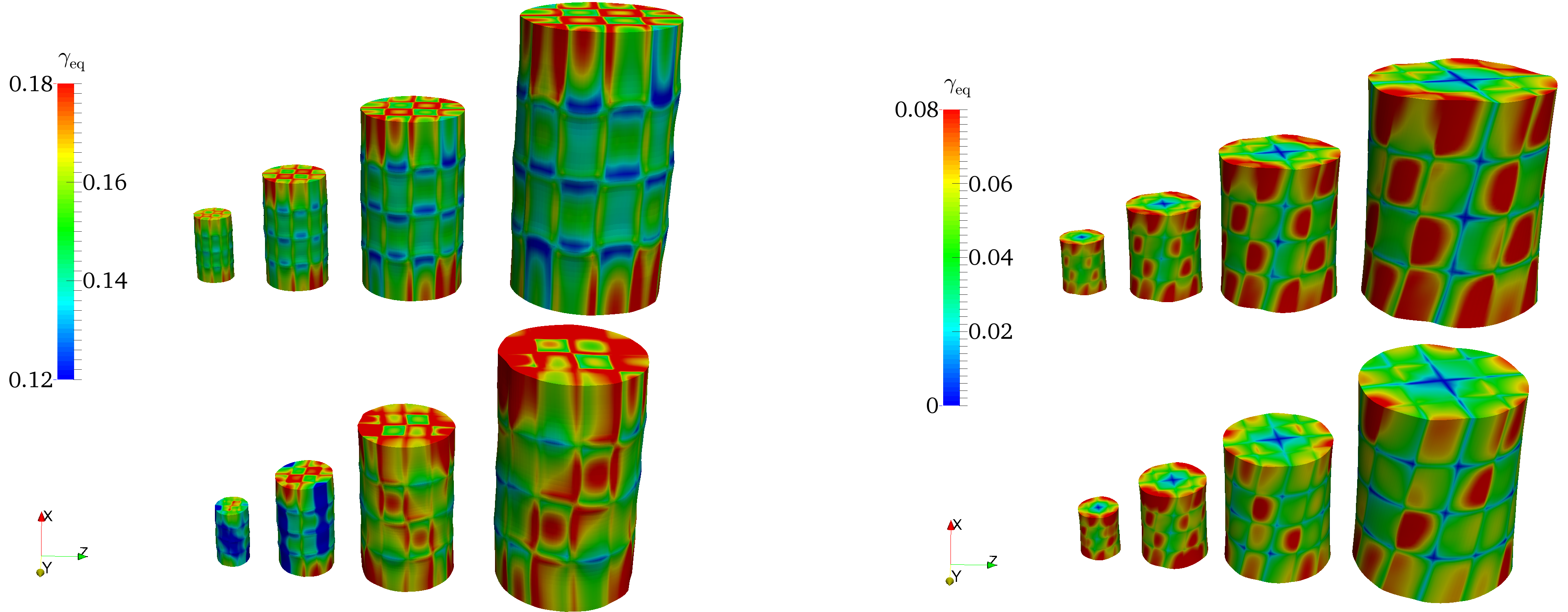}\\
 \begin{minipage}{.45\textwidth}
  \hspace{38mm}{\bf (a)}
 \end{minipage}
 \begin{minipage}{.45\textwidth}
  \hspace{40mm}{\bf (b)}
\end{minipage}
\caption{Field distributions of equivalent plastic strain for simplified grain aggregates using equal cross-section area shares of \f{\langle111\rangle}- and \f{\langle100\rangle}-orientations (top) and using experimentally determined, simplified cross-section area shares of \f{\langle111\rangle}- and \f{\langle100\rangle}-orientations (bottom). {\bf a} Tensile test simulations, {\bf b} Torsion test simulations. The deformation of all results has been geometrically scaled by a factor of five with regard to the displacement.}
\label{fig:visualization_zeta}
\end{center}
\end{figure*}

In order to investigate the influence of the specific texture in each microwire, simulations considering the trend towards higher \f{\langle100\rangle}-texture for smaller wire diameters are carried out. Therefore, the experimental results for the individual texture obtained from the approach in \figref{fig:exp_fig} are used (see Table~\ref{tab:exps} for experimental data and \figref{fig:grains}b-e for the grain aggregates). The model parameters are fitted to the experimental tensile response, again (see \figref{fig:size_70_30}a).
The resulting set of parameters is listed in Table~\ref{tab:zug_k_kG100_70_30_NEU}.
\begin{table}[htbp]
\centering
\renewcommand{\arraystretch}{1.2}
\caption{Model parameters calibrated to tensile data using the experimentally determined, simplified cross-section area shares of \f{\langle111\rangle}- and \f{\langle100\rangle}-orientations}
\begin{tabular}{ccccc}
\hline
\f{D_{\rm sim}}&\f{\tau_0^{\rm C}}&\f{\Xi_{0}^{\rm C}}&\f{\Theta}&\f{\tau_\infty^{\rm C}}\\
($\upmu$m)&(MPa)&(Nm$^{-1}$)&(MPa)&(MPa)\\\hline
15&1&13&143&21\\
25&1&19&135&28\\
40&1&25&143&37\\
60&1&37&145&35\\
\hline
\end{tabular}
\label{tab:zug_k_kG100_70_30_NEU}
\end{table}
Using these parameters in the torsion test simulations, the results in \figref{fig:size_70_30}b are obtained. A pronounced size effect of the grain aggregate simulations is observable, although the experimental tensile responses, showing an inverse size effect, have been used to obtain the model parameters. The magnitude of the size-effect under torsion loading, however, is underestimated. \revM{It should also be noted that the magnitude of the modeled size effect under torsion becomes smaller for increasing deviations of the texture of the two thicker microwires from the simplified approach taken for the grain aggregates (compare the two curves of the thicker grain aggregates in \figref{fig:size_70_30} to the respective curves \figref{fig:size_50_50}).}\\
The distributions of the equivalent plastic strain~\f{\gaq} for tensile and torsion loading with and without consideration of the simplified microwire texture are depicted in \figref{fig:visualization_zeta}a-b. Under both loading conditions, the field distributions among the grain aggregates are very similar if the grain orientations are distributed equally. If the simplified microwire textures are considered, however, different field distributions of the equivalent plastic strain are obtained for the investigated aggregates of different diameters. \rtwoC{It is remarked that the locally developing high values of~\f{\gaq} at the corners of the grains close to the GBs (see, for example, the top surface of the largest grain aggregate in \figref{fig:visualization_zeta}b~(bottom)) result from the activation of favorably oriented slip systems in combination with the impediment of plasticity at the GB nodes due to the GB yield strength. In the ideally aligned $\langle100\rangle$-orientation, the slip systems are activated such that a fourfold-symmetry of~\f{\gaq} would be obtained for a single-crystalline torsion test simulation (see, e.g., Fig.~5 in \cite{ziemann2015deformation}). The spatial positions of the maxima of~\f{\gaq} result from the activation of the slip systems, keeping the $\langle100\rangle$-orientation in mind. Therefore, higher values of~\f{\gaq} are observable close to the corners of some of the grains. A similar localization would also occur, if the GB was modeled using a curved shape, there.}\\
\revM{From Table~\ref{tab:zug_k_kG100_50_50_NEU} and Table~\ref{tab:zug_k_kG100_70_30_NEU} it is evident that different values of the grain boundary yield strength~\f{\Xi^{\rm C}_0} have to be used in order to fit the experimental results. This is due to the computational limitation regarding the simulated average grain sizes. If all experimental average grain sizes could be matched exactly in the simulations, a wire-diameter-independent value of the GB yield strength would be expected. The different values of~\f{\Xi^{\rm C}_0}, therefore, obviously compensate for the deviations in the modeling of the grain size.}
\subsubsection{Gradient plasticity simulations considering the simplified texture of the microwires and scaling of parameters}\label{sec:scaled}
In the experiments, an increase in the overall yield strength is observed comparing the torsion test results to the tensile test results of the microwires. Therefore, the parameters used for torsion loading are scaled accordingly. It is investigated how this affects the size-effect prediction of the model. Since it appears that the initial overall yield strength is significantly underestimated (see \figref{fig:size_70_30}b) by keeping the initial yield stress of the slip systems as well as the grain boundary yield strength identical under torsion loading, these two model parameters (which constitute the overall yield strength, e.g., \f{R_{p0.2}} in the model response) are altered.\\
\mw{The overall initial shear stresses are estimated based on the experimentally determined individual proof stresses \f{\tau_{0.2}^{\rm te}} and \f{\tau_{0.2}^{\rm to}}, respectively (see  Table~\ref{tab:zug_vs_torsion})}.
\begin{table}[htbp]
\centering
\renewcommand{\arraystretch}{1.2}
\caption{Yield strength scaling: the maximum shear stresses at the onset of yielding in the experiments ($^1$: from the experiment corresponding to wire diameter~$D$ in~\cite{chen2015size}) are used to determine scaling factors for the model parameters under torsion loading. \f{\tau_{0.2}^{\rm te}} obtained from experimental engineering stress, \f{\tau_{0.2}^{\rm te}=0.5\sigma_{{\rm eng},0.2}}}
\label{tab:zug_vs_torsion}
\begin{tabular}{cccc}
\hline
Wire&tensile shear&torsion shear&Shear stress\\
\f{D_{\rm sim}}&stress$^1$ \f{\tau_{0.2}^{\rm te}}&stress$^1$ \f{\tau_{r=R,0.2}^{\rm to}}&ratio \f{s=}\\
($\upmu$m)&(MPa)&(MPa)&\f{\tau_{r=R,0.2}^{\rm to}/\tau_{0.2}^{\rm te}}\\\hline
15&15.8&54.3&3.4\\
25&16.4&39.5&2.4\\
40&14.4&31.7&2.2\\
60&13.9&22.7&1.6\\
\hline
\end{tabular}
\end{table}
The scaling factors from the initial overall yield strengths under both loading conditions in the experiments are then used to scale the obtained model parameters from the tensile loading to model parameters for the torsion loading. The resulting parameters are shown in Table~\ref{tab:zug_k_kG100_30_70_NEU_c_t}. It is remarked that the saturation stresses~\f{\tau_\infty^{\rm C}} of the slip systems are adjusted such that the work-hardening behavior of the microwires is kept the same, compared to the cases without scaling the initial yield stresses~\f{\tau_0^{\rm C}}.
\begin{table}[htbp]
\centering
\renewcommand{\arraystretch}{1.2}
\caption{Torsion test simulation parameters, \f{\tau_0^{\rm C}} and \f{\Xi_{0}^{\rm C}} from Table (T) scaled by~\f{s} from Table~\ref{tab:zug_vs_torsion}, \f{\tau_{\infty,s}^{\rm C}} obtained by keeping hardening behavior (i.e., \f{\Delta\tau^{\rm C}=\tau_{\infty}^{\rm C}-\tau_0^{\rm C}}) identical to the tensile test}
\label{tab:zug_k_kG100_30_70_NEU_c_t}
\begin{tabular}{cccccc}
\hline
\f{D_{\rm sim}}&\f{s\tau_0^{\rm C}}&\f{s\Xi_{0}^{\rm C}}&\f{\Theta}&\f{\tau_{\infty,s}^{\rm C}=}&T\\
&&&&\f{s\tau_0^{\rm C}+\Delta\tau^{\rm C}}&\\
($\upmu$m)&(MPa)&(Nm$^{-1}$)&(MPa)&(MPa)&\\\hline
15&3.4&44&143&23.4&\ref{tab:zug_k_kG100_70_30_NEU}\\
25&2.4&46&135&29.4&\ref{tab:zug_k_kG100_70_30_NEU}\\
40&2.2&55&143&38.2&\ref{tab:zug_k_kG100_70_30_NEU}\\
60&1.6&59&145&35.6&\ref{tab:zug_k_kG100_70_30_NEU}\\[1.5ex]
40&2.2&64&144&41.2&\ref{tab:zug_k_kG100_50_50_NEU}\\
60&1.6&69&155&39.6&\ref{tab:zug_k_kG100_50_50_NEU}\\
\hline
\end{tabular}
\end{table}
\begin{figure*}[htbp]
\begin{center}
\includegraphics[width=0.95\linewidth]{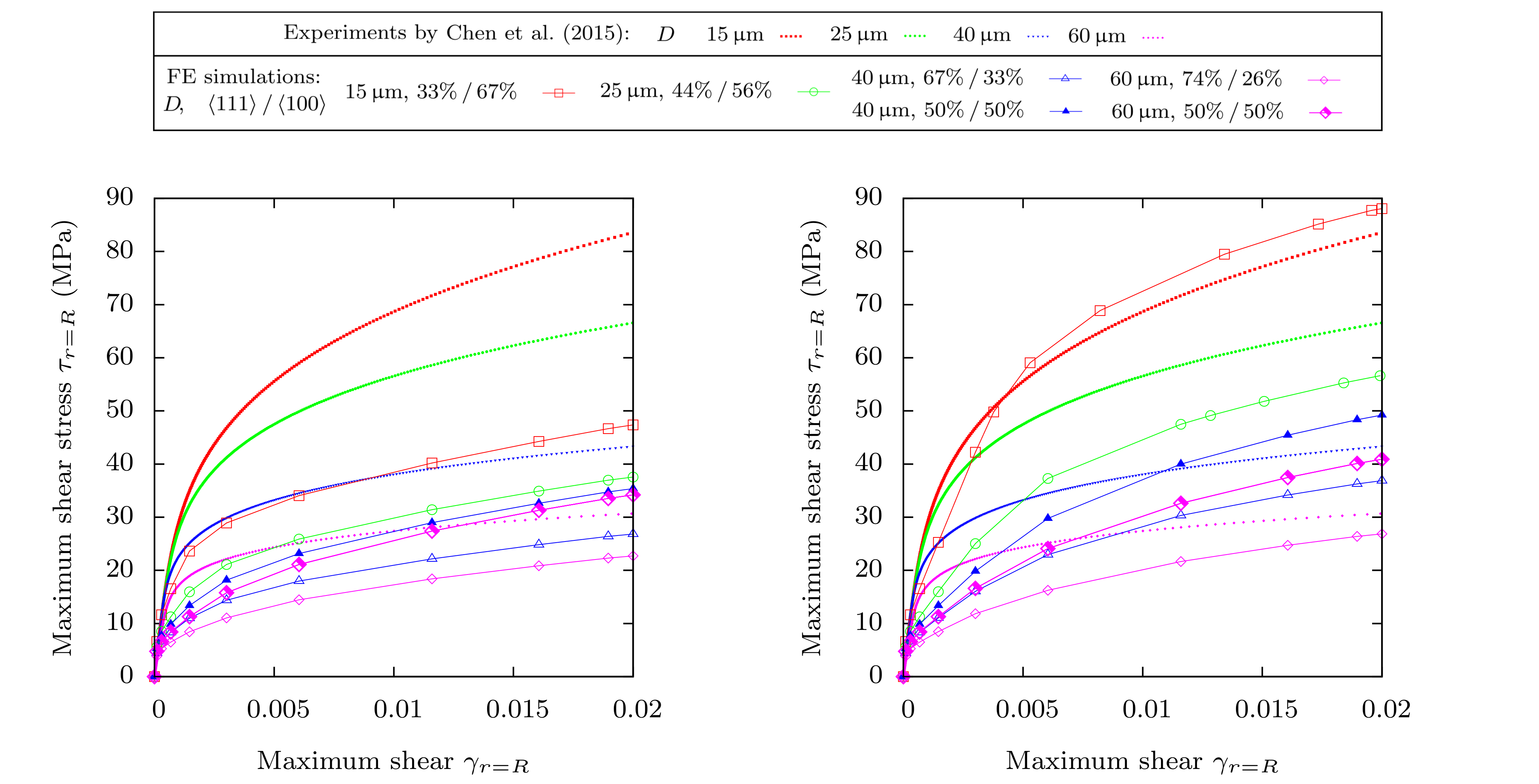}\\
 \begin{minipage}{.45\textwidth}
  \hspace{33mm}{\bf (a)}
 \end{minipage}
 \begin{minipage}{.45\textwidth}
  \hspace{40mm}{\bf (b)}
\end{minipage}
\caption{Maximum shear stress vs.\ maximum shear of torsion-loaded simplified grain aggregates in comparison to experimentally observed oligocrystalline microwire responses by~\cite{chen2015size}. {\bf a} using scaled initial yield stress~\f{s\tau_0^{\rm C}} of the slip systems, {\bf b} using scaled initial yield stress~\f{s\tau_0^{\rm C}} of the slip systems and scaled yield strength~\f{s\Xi_{0}^{\rm C}} of the GBs. For each microwire diameter, the experimentally determined, simplified cross-section area shares of \f{\langle111\rangle}- and \f{\langle100\rangle}-orientations are used in the FE simulations. For the two thicker microwires, additional FE simulations using scaled model parameters and equal cross-section area shares of \f{\langle111\rangle}- and \f{\langle100\rangle}-orientations are depicted}
\label{fig:size_70_30_scaled}
\end{center}
\end{figure*}
At first, only the initial yield stress of the slip systems is scaled (see Table~\ref{tab:zug_k_kG100_30_70_NEU_c_t}, second column) but the yield strength of the GBs is kept identical to the parameters obtained from the tensile loading (see Table~\ref{tab:zug_k_kG100_50_50_NEU} and Table~\ref{tab:zug_k_kG100_70_30_NEU}). The torsion test simulation results obtained by this approach are depicted in \figref{fig:size_70_30_scaled}a. Although the observed size effect magnitude in the simulations is slightly larger than using the previous initial yield stresses, the employed scaling of the slip system yield stresses is not sufficient to account for the size effect. For the two thicker microwires, the grain aggregates using equal cross-section area shares of both orientations show stiffer responses than the ones using the experimentally determined, simplified area shares of the orientations. The mechanical response of the microwire with \f{D=60\,\upmu}m is captured better by using equal cross-section area shares of both orientations. However, for the microwire with \f{D=40\,\upmu}m, the mechanical response is still underestimated using equal shares.\\
Since the yield strength of the GBs contributes to the overall yield strength of the grain aggregates in the simulations, this parameter is also scaled, subsequently (see Table~\ref{tab:zug_k_kG100_30_70_NEU_c_t}, third column). The results are depicted in \figref{fig:size_70_30_scaled}b. It is observable that the magnitude of the size effect is captured much better by scaling both the initial slip system yield stress and the grain boundary yield strength. In addition, the mechanical responses of the simulations are much closer to the experimental results for all microwires. It is also notable that the experimental results of the thicker microwires are both in-between the respective two simulated cases of the orientation distributions in the cross sections.\\
The field distributions of the equivalent plastic strain~\f{\gaq} of the results with scaled initial yield stress of the slip systems are depicted in \figref{fig:visualization_zeta_scaled}a, and the distributions with additionally scaled GB yield strength are depicted in \figref{fig:visualization_zeta_scaled}b. If only the initial slip system yield stress is scaled, the magnitude of the field distributions is only slightly reduced (see \figref{fig:visualization_zeta_scaled}a in comparison to \figref{fig:visualization_zeta}b). However, if the GB yield strength is additionally scaled, the GBs respond significantly stiffer and less equivalent plastic strain is observable, there (see arrows in \figref{fig:visualization_zeta_scaled}b).
\begin{figure*}[htbp]
\begin{center}
 \includegraphics[width=0.95\linewidth]{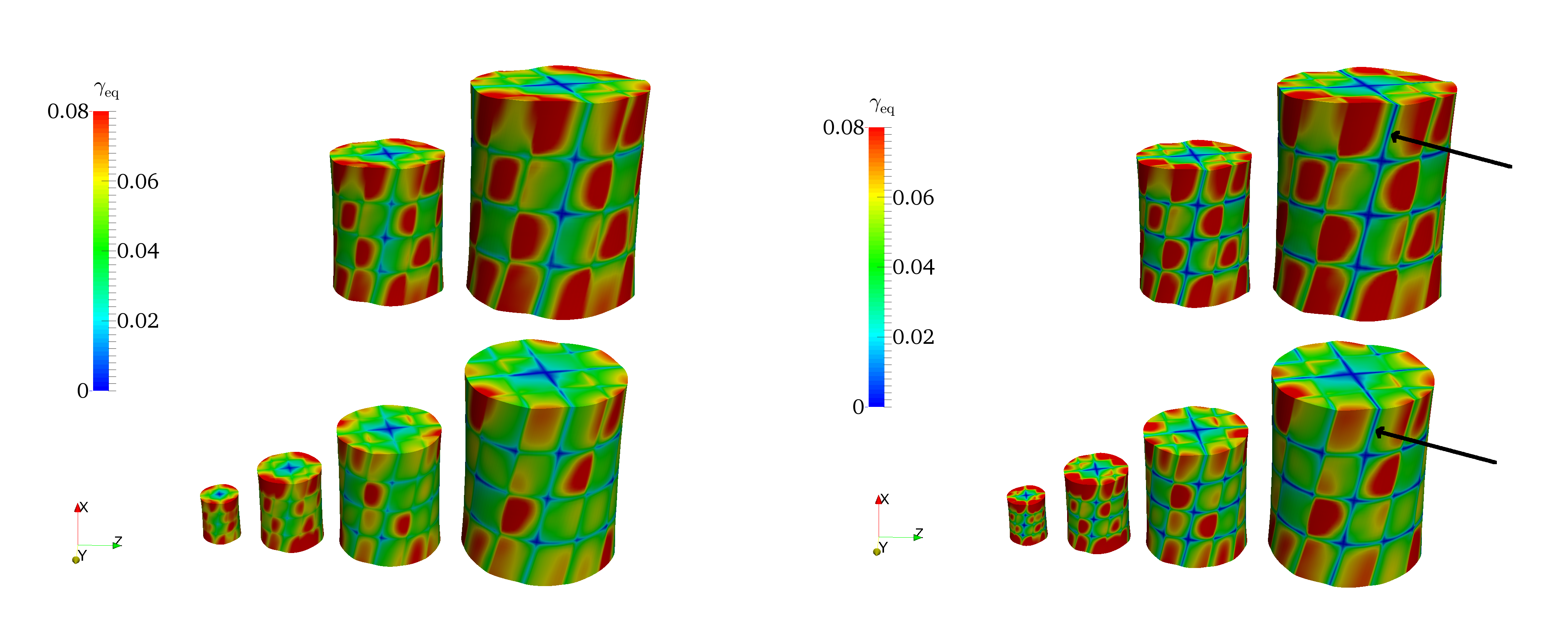}\\
 \begin{minipage}{.45\textwidth}
  \hspace{38mm}{\bf (a)}
 \end{minipage}
 \begin{minipage}{.45\textwidth}
  \hspace{40mm}{\bf (b)}
\end{minipage}
\caption{Torsion test simulations: field distributions of equivalent plastic strain for simplified grain aggregates using equal cross-section area shares of \f{\langle111\rangle}- and \f{\langle100\rangle}-orientations (top) and using experimentally determined, simplified cross-section area shares of \f{\langle111\rangle}- and \f{\langle100\rangle}-orientations (bottom). {\bf a} the initial yield strength of the slip systems has been scaled, {\bf b} the initial yield strength of the slip systems and the grain boundary yield strength have been scaled. The deformation of all results has been geometrically scaled by a factor of five with regard to the displacement}
\label{fig:visualization_zeta_scaled}
\end{center}
\end{figure*}
This can be inspected more closely by evaluating the average equivalent plastic strain over circular line segments on the top surface (i.e., the surface at max.\ $x$-coordinate), see \figref{fig:zeta_vs_r}a. Comparing these averaged distributions for the first five time steps shows that the scaling of the initial slip system yield stress slightly reduces the plastification in radial direction (see II in \figref{fig:zeta_vs_r}b in comparison to I). The scaling of the GB yield strength, however, leads to pronounced gradients in the equivalent plastic strain field distribution close to the GBs (see III in \figref{fig:zeta_vs_r}b and the arrow-indicator at \f{r=4\,\upmu}m, near the GB).
\begin{figure*}[htbp]
\begin{center}
 \includegraphics[width=0.9\linewidth]{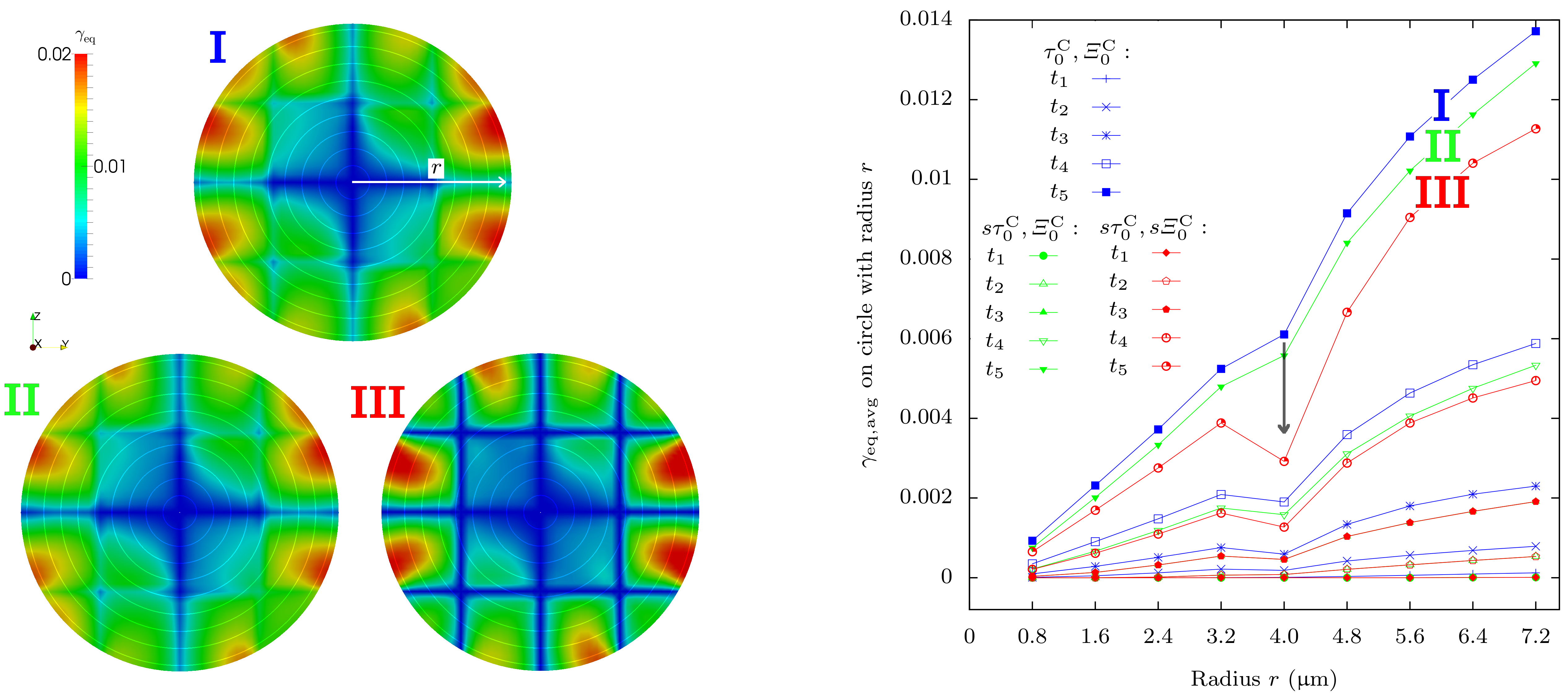}\\
 \begin{minipage}{.45\textwidth}
  \hspace{30mm}{\bf (a)}
 \end{minipage}
 \begin{minipage}{.45\textwidth}
  \hspace{48mm}{\bf (b)}
\end{minipage}
\caption{{\bf a} Torsion test simulations using experimentally determined, simplified cross-section area shares of \f{\langle111\rangle}- and \f{\langle100\rangle}-orientations: field distributions of equivalent plastic strain on the top surface of $15\,\upmu$m-grain-aggregate at the end of the fifth time-step for the cases of unscaled (I) and scaled parameters (II/III). The used line segments for obtaining the plots in {\bf b} are visualized. {\bf b} Average equivalent plastic strain evolution during the first five time steps on the corresponding circular line segments for the cases of unscaled (I), and scaled model parameters (II/III)}
\label{fig:zeta_vs_r}
\end{center}
\end{figure*}
\section{Discussion of results}
Aggregates of grains with simplified shape and orientation distributions have been used in gradient plasticity simulations to model the mechanical response of oligocrystalline microwires. \rtwoC{Considering that the (non-destructive) experimental 3D-characterization of grain morphologies has made significant progress during recent years (see, e.g., \cite{cabus2014influence,hounkpati2014influence,toda2016diffraction} and the review by \cite{maire2014quantitative}), it would be desirable to perform simulations using the actual morphology of all grains of the microwires, in order to have a representation of the microstructure in the simulations being closer to the experiment. For such an approach, the FE-meshes need to be substantially refined, for example, when modeling smaller grains of non-spheroidal or non-cuboid shape. Therefore, this would require further numerical improvement of the computational approach. Within the present study, however, only grains of cuboid shape could be used.} The average grain size of the microwires is considered in the simulations but the variations of grain size for different grains within the cross-sections are not (Table~\ref{tab:exps} and Table~\ref{tab:disc}). Due to computational limitations, only aggregates up to 64~grains could be used. \mww{This leads to remaining smaller deviations for two and larger deviations for the other two microwires in terms of the average grain size employed}.\\
Using equally distributed, ideally oriented, \f{\langle100\rangle}- and \f{\langle111\rangle}-grains, the experimentally observed tensile response of microwires by~\cite{chen2015size} could be fitted with the employed gradient plasticity model (\figref{fig:size_50_50}a). The torsion response obtained in simulations with identical parameters and texture as in the tensile response, however, shows similar behavior as the tensile response, and does not match the classic size effect observed in the experiments for this loading (\figref{fig:size_50_50}b).\\
If the (simplified) texture of the microwires is considered in the grain aggregate simulations, the tensile response can be fitted (\figref{fig:size_70_30}a). Then, a pronounced size effect is observable under torsion loading (\figref{fig:size_70_30}b). The arising differences in the torsion responses regarding the texture can be attributed to the different orientations of slip systems. The ideal \f{\langle100\rangle}-orientation has more slip systems that are favorably oriented for an activation under the investigated tensile loading compared to the ideal \f{\langle111\rangle}-orientation. This behavior is reversed under torsion loading. Therefore, grain aggregates with a higher cross-section share of \f{\langle100\rangle}-orientation respond significantly stiffer under torsion loading than grain aggregates with a higher \f{\langle111\rangle}-orientation cross-section share. \revM{It should be noted that the magnitude of the size effect in the model varies, depending on the chosen texture of the two larger aggregates (compare, for example, the two curves of the thickest grain aggregate in \figref{fig:size_70_30_scaled}a). Considering that, for example, the largest microwire (see, for example, \figref{fig:exp_fig}f) shows larger area shares with higher misorientations from the two modeled ideal orientations, one would expect the texture-contribution to the size effect to be smaller than it is obtained by using the idealized texture from \figref{fig:grains}e.}\\
The fitted model parameters are not unique for both cases. In parts, they show similar values and overall they show the same tendencies, e.g., the increase in the slip system saturation stress with increasing wire diameter (Table~\ref{tab:zug_k_kG100_50_50_NEU} and Table~\ref{tab:zug_k_kG100_70_30_NEU}). This increase saturates for the two largest microwires.\\
\revC{In the initial development of the present model, the GB yield strength has been intended to be a material constant. However, using a unique value of the GB yield strength for all investigated microwire diameters is not applicable. On this account, it should be noted that the ratio of average grain size to wire diameter, in the experiments, is not unique for different wire diameters. For the simulated grain aggregates, however, the same ratio has been used for all discretizations. Additionally, in the experiments, the microstructure of the microwires has been annealed by different heat treatments for different wire diameters. More refined grain aggregates would be desirable to be used in the simulations but the arising computational costs appear to be prohibitive at the current time. Therefore, the GB yield strength is, e.g., chosen substantially larger for the biggest aggregate to compensate for the difference in the grain size. Summarizing, it appears to be necessary to adapt the values of the GB parameter in the model to the specific microwires when the grain size can not be modeled exactly.}\\
\rtwoC{Due to the fact that all considered microwires in this work exhibit comparably low shares of low-angle GBs (on average \f{<10\%} for angles \f{<10^\circ}), and that a model with a non-orientation dependent GB yield strength is used, effects of varying in-plane GB misorientations are not investigated. Instead, it is focused on the effects stemming from the crystal-orientation dependent activation of slip systems for the two distinct orientations considered in the modeling approach.}\\
The consideration of the simplified microwire textures enables to capture both the inverse size effect under tensile loading and the classic size effect under torsion loading. However, the magnitude of the size effect under torsion loading is underestimated (\figref{fig:size_70_30}b). \mww{Scaling the initial yield stress of the slip systems by taking into account the experimentally determined 0.2-proof-stresses improves the simulation results, but not enough} to obtain a good estimate of the size effect magnitude (\figref{fig:size_70_30_scaled}a). If, additionally, the GB yield strength is scaled, a better approximation of the size effect magnitude is obtained (\figref{fig:size_70_30_scaled}b). In parallel, the simulated equivalent plastic strain gradients close to the GBs are increased by this procedure (\figref{fig:zeta_vs_r}b). Such gradients in the model serve as an approximation of the inhomogenity of the plastic deformation. The GND-densities calculated from the experiments show pronounced variations due to the microstructural characteristics (see, e.g., Fig.~4.30 in~\cite{chen2013deformation}, however, for larger deformations than in the present work). Thus, the simulation results with larger gradients seem to be closer to the experiments. A more detailed assessment requires the determination of GND-densities in many cross-sections of all considered microwires for smaller overall deformations (than in~\cite{chen2013deformation}).\\
From a materials science point of view, it appears questionable that a different GB yield strength in the torsion test simulations results in a better agreement with the experiments than using the values as obtained from the tensile test fits. However, the initial distribution of dislocations has not been considered in the simulations. Therefore, possibly differing activation stresses within the dislocation distributions of the same wires under different loading conditions are not modeled. This might be caused by the activation of dislocations close to the surface of the microwires under torsion loading. The scaling of initial slip system stress and GB yield strength for the torsion loading simulations can only give an estimate of the mechanical response. \rtwoC{The used model for GB yielding does not take into account the different orientations of the grains. In general, however, one would expect the GB yield strength to depend on the crystallographic orientations of the adjacent grains and of the GB itself, i.e., on the specific type of the GB. Several approaches to consider the crystallographic misorientation have been proposed in the literature, see \cite{bayerschen2016review} for a recent review.}\\
In the overall torsion response of simulations and experiments, deviations remain for the microwire with \f{D=25\,\upmu}m (\figref{fig:size_70_30_scaled}b). These deviations could be reduced by employing a smaller grain size in the simulations that is closer to the experimentally determined average grain size. For the largest microwire, it would also be desirable to perform simulations with a refined grain size, although the influence of the grain size for the larger microwires is not expected to be as crucial as it is for smaller microwires.\\
The texture of the microwires has been considered in the simulations by a simplified approach (\figref{fig:exp_fig}b). A refined approach would require the experimental characterization of many cross sections to obtain detailed results of the variations of texture with regard to the deviations from the ideal \f{\langle100\rangle}- and \f{\langle111\rangle}-orientations. The approach of a simplified texture characterization appears justifiable for the two smaller microwires (\figref{fig:exp_fig}ad) that show less cross-section shares with orientations deviating from the two considered ideal orientations. For the two thicker microwires (\figref{fig:exp_fig}ef), however, the examined cross-sections show more area shares with orientations deviating from the ideal orientations investigated. Therefore, two texture cases (for \f{D=40\,\upmu}m with \figref{fig:grains}a and \figref{fig:grains}d, for \f{D=60\,\upmu}m with \figref{fig:grains}a and \figref{fig:grains}e, respectively) have been used in the simulations to obtain an estimate of the mechanical responses. The mechanical response of the microwires in the experiment is in between the responses of the two simulated texture cases (\figref{fig:size_70_30_scaled}ab) for both diameters. Thus, the real texture of the microwires could be in between the two idealized cases simulated.\\
Regarding the employed gradient plasticity model, several simplifications are addressed, in the following. First of all, an equivalent plastic strain and its gradient are used instead of considering all plastic slips and their gradients. Physically richer models incorporate all plastic slips or dislocation densities, e.g.,~\cite{Gurtin2007,bardella2013latent} as additional degrees of freedom. Using the equivalent plastic strain and its gradient leads to a reduction of the defect stress modeling since only one higher-order stress, work-conjugate to the gradient of \f{\gaq}, is modeled. Considering all gradients would allow to include backstresses related to the individual slip systems but the model merit of manageable computational times would be lost. In addition, it has been assumed that the equivalent plastic strain measure is continuous across GBs. The plastic slips of individual slip systems are, however, not necessarily continuous. Although this simplification could be justifiable in certain cases, it is not expected to be of general validity~\cite{bayerschen2015equivalent}.\\ Furthermore, the equivalent plastic strain has been set to vanish throughout the simulation volume, in the beginning of the simulation. This appears to be reasonable since the annealing process produces oligocrystalline microwires with comparably low initial dislocation density content. It would, however, be desirable to consider the initial dislocation density distribution. Detailed data of the dislocation density distributions within all grains and in many cross sections would need to be obtained for meaningful non-zero choices of the initial equivalent plastic strain distribution. The effort for this appears to be prohibitive at the current time.\\
As it is also commonly done in the literature, the defect energy related to the gradients in the equivalent plastic strain (see Table~\ref{tab:FreeEnergyDensities}) has been taken to be of quadratic form. However, other approaches (e.g., linear) have been proposed in the literature as well. Different equivalent plastic strain distributions would be obtained with non-quadratic defect energies. This includes steeper gradients close to GBs for, e.g., a linear defect energy compared to a quadratic one. The defect energy also influences the magnitude of size effects. Therefore, it could be interesting to additionally consider the influence of the defect energy formulation in future works.\\
\revC{The parameter of the defect energy has been fitted to the experimental tensile responses. Since a simplistic quadratic defect energy is used in the current model, a direct connection to the microstructural internal length is not necessarily given. Recent works suggest that the free energy of dislocations should in fact be non-quadratic, e.g., \cite{kooiman2015microscopically,kooiman2016free}. Although the internal length scale is, in general, not expected to be constant, the accordance with the experimental tensile test results is given when using the same value of the defect energy parameter for all grain aggregates in the present approach. From a physical point of view, the internal length scale is determined by the microstructural characteristics of the specimen such as grain size, specimen size, dislocation spacing and dislocation source length, see, e.g., \cite{zhang2014}.} \revC{The obtained value for the internal length scale of~\f{l_{\rm int,avg}=0.1\,\upmu}m is, however, at least of the same order as, for example, the value of the internal length scale of the gradient plasticity model \cite{aifantis2004interfacial,aifantis2005role} which was determined in nanoindentation studies close to a grain boundary~\cite{aifantis2006interfaces}. There, the interpretation was given that this length was the distance over which 90\% of the dislocations are piled up.}\\ \mww{The used phenomenological work-hardening formulation of Voce-type can be derived by combining the Taylor-relation with the Kocks-Mecking model, for monotonic loading processes with constant local slip rate \cite{mecking2001work,kocks2003physics}.} Since the plastic slip parameters of the employed gradient plasticity model are non-decreasing by definition, this approach is only applicable to monotonic loading processes, and Voce-type hardening can be used.\\
\mw{Although the actual model does not allow to predict the torsional response of Au microwires of different thickness on the basis of the results from related tensile tests exactly, it is clearly shown that all different microstructural features have to be taken into account when modeling the mechanical behavior of small structures under different load conditions. In the present case it is particularly demonstrated, that the size effect in torsion, determined in \cite{chen2015size}, is significantly affected by textural differences. If all the wires would reveal comparable textures, the difference in strength of the smaller wires, compared to the bigger ones, would be much smaller. In fact, diameter depending textural differences of heat-treated coarse-grained microwires, tested in tension and torsion, are obviously not only occurring in samples investigated by Chen et al. When viewing the results of tensile tests on Cu microwires from different corresponding publications, pronounced differences in the hardening behavior of the microwires of different thickness are also observable (e.g., Fig.~1 in \cite{liu2012size}, Fig. 2 in \cite{liu2013toward}, and, at least to some extent, in Fig.~5 in \cite{fleck1994strain}). For lower initial dislocation densities and less strong varying average grain sizes one may assume, that these differences are also mainly caused by textural variations.}
\section{Conclusion}
\mww{Three-dimensional finite element simulations were performed to model the deformation behavior of oligocrystalline gold microwires in both tension and torsion loading, using simplified grain aggregates.} A gradient plasticity framework based on an equivalent plastic strain and its gradient is used. It is demonstrated that the consideration of the texture of the microwires, although simplified here, is crucial for modeling the contrary size effects under different loading conditions. The overall inverse size effect observed in the tensile experiments of Chen et al.\ (2015, {\it Acta Mater.} {\bf 87}, 78-85) is fitted with the model. A pronounced size effect is observed in the same work under torsion loading, which could be reproduced to some extent. Based on a scaling of the yield strengths, informed by the comparison of experimental responses under both loading conditions, an estimate of the mechanical response under torsion loading is obtained.\\

{\bf\flushleft Conflicts of interest}\\
The authors confirm that no conflicts of interest arise with regard to the research leading to this paper, nor with publication of this work.\\

{\bf\flushleft Compliance with ethical standards}\\
No ethical standards were violated during the performed experimental and computational investigations.\\



\noindent{\bf Acknowledgements}\\
The authors acknowledge the support rendered by the German Research Foundation (DFG) under the Grants {BO1466/5-{\color{black}2}} and {GR3677/2-{\color{black}2}}. The funded projects "Dislocation based Gradient Plasticity Theory" and "Experimental Characterization of Micro Plasticity and Dislocation Microstructure" are part of the DFG Research Group 1650 "Dislocation based Plasticity".
\bibliographystyle{myunsrt}
\bibliography{lit_abb}
\end{document}